\documentclass[aps,showpacs]{revtex4}%
\usepackage{graphicx}
\usepackage{amsmath}
\usepackage{amsfonts}
\usepackage{amssymb}
\usepackage{amsmath}
\usepackage{latexsym,amssymb}
\usepackage{epsfig}
\usepackage{graphicx}
\usepackage{bm}%
\begin{document}
\title{Phantom stars and topology change}
\author{Andrew DeBenedictis}
\email{adebened@sfu.ca}
\affiliation{Pacific Institute for the Mathematical Sciences, Simon Fraser University Site}
\affiliation{Department of Physics, Simon Fraser University Burnaby, British Columbia, V5A
IS6, Canada}
\author{Remo Garattini}
\email{Remo.Garattini@unibg.it}
\affiliation{Universit\`{a} degli Studi di Bergamo, Facolt\`{a} di Ingegneria, Viale
Marconi 5, 24044 Dalmine (Bergamo) ITALY}
\affiliation{INFN - sezione di Milano, Via Celoria 16, Milan, Italy.}
\author{Francisco S. N. Lobo}
\email{francisco.lobo@port.ac.uk}
\affiliation{Institute of Cosmology \& Gravitation, University of Portsmouth, Portsmouth
PO1 2EG, UK}
\affiliation{Centro de Astronomia e Astrof\'{\i}sica da Universidade de Lisboa, Campo
Grande, Ed. C8 1749-016 Lisboa, Portugal}
\date{\today}

\begin{abstract}
In this work, we consider time-dependent dark energy star models, with an
evolving parameter $\omega$ crossing the phantom divide, $\omega=-1$. Once in
the phantom regime, the null energy condition is violated, which physically
implies that the negative radial pressure exceeds the energy density.
Therefore, an enormous negative pressure in the center may, in principle,
imply a topology change, consequently opening up a tunnel and converting the
dark energy star into a wormhole. The criteria for this topology change are
discussed and, in particular, we consider a Casimir energy approach involving
quasi-local energy difference calculations that may reflect or measure the
occurrence of a topology change. We denote these exotic geometries consisting
of dark energy stars (in the phantom regime) and phantom wormholes as
\textit{phantom stars}. The final product of this topological change, namely,
phantom wormholes, have far-reaching physical and cosmological implications,
as in addition to being used for interstellar shortcuts, an absurdly advanced
civilization may manipulate these geometries to induce closed timelike curves,
consequently violating causality.

\end{abstract}

\maketitle


\section{Introduction}


Recent high-precision observational data have confirmed that the
Universe is undergoing a phase of accelerated expansion
\cite{expansion}. Several candidates, responsible for this
expansion, have been proposed in the literature, in particular,
dark energy models (see Ref. \cite{Copeland} for a review) and
modified gravity (e.g., see Refs. \cite{modgravity} for recent
reviews). In particular, the former models are fundamental
candidates, in which a simple way to parameterize the dark energy
is by an equation of state of the form $\omega\equiv p/\rho$,
where $p$ is the spatially homogeneous pressure and $\rho$ is the
dark energy density. A value of $\omega<-1/3$ is required for
cosmic expansion, and $\omega=-1$ corresponds to a cosmological
constant. A specific exotic form of dark energy denoted phantom
energy, with $\omega<-1$, has also been proposed \cite{Caldwell},
and possesses peculiar properties, such as the violation of the
null energy condition (NEC) and the energy density increases to
infinity in a finite time \cite{Caldwell}, at which point the size
of the Universe blows up in a finite time, which is known as the
Big Rip. In this context, the violation of the NEC presents us
with a natural scenario for the existence of traversable
wormholes, and indeed it has been shown that these exotic
geometries can be supported by phantom energy
\cite{Sushkov,Lobo-phantom}. It is also interesting to note that
recent fits to supernovae, CMB and weak gravitational lensing data
probably favor an evolving equation of state, with the parameter
crossing the phantom divide $\omega=-1$ \cite{Vikman}.

Despite the fact that the dark energy equation of state represents
a spatially homogeneous cosmic fluid and is assumed not to
cluster, it is possible that inhomogeneities may arise due to
gravitational instabilities. More precisely, although the equation
of state leading to the acceleration of the Universe on large
scales is an average equation of state corresponding to a
background fluid, it is possible that dark energy condensates may
possibly originate from density fluctuations in the cosmological
background, resulting in the nucleation through the respective
density perturbations. Despite the fact that once in the dark
energy regime the material system becomes gravitationally
repulsive, we may consider the possibility of the formation of a
matter system that originally obeys all the energy conditions.
Cosmological observations do not rule out, and in some studies
favor, an evolving equation of state for the dark energy. It is
therefore quite possible that what we know as dark energy today
has evolved from a more benign fluid. An over-density of this
fluid could in principle commence a collapse into a star which.
Such a model is presented in section \ref{sec:IIIC}. We also point
out that even in the case of a dark-energy fluid, there is no
definite resolution to the debate of clustering scales. This is
mainly due to non-linearity, especially in the vein of dark energy
interacting with ordinary fluids. It may also be possible to glean
some information on the cosmological dark matter by studying
certain properties of such gravitational condensates. (See
\cite{ref:aandaandijtp} and references therein for comments on
these issues.) In this context, a number of inhomogeneous
solutions have been the object of analysis, such as the phantom
wormholes \cite{Sushkov,Lobo-phantom} mentioned above, dark energy
stars \cite{Lobo:2005uf}, and other structures such as condensates
supported by the generalized Chaplygin gas \cite{Chapgas} which
possibly arise from density fluctuations in the generalized
Chaplygin gas background, and condensed
structures supported by the van der Waals equation of state \cite{Lobo:2006ue}%
. In a recent paper \cite{Dzhunushaliev:2008bq}, it was also shown that the
$4D$ Einstein-Klein-Gordon equations with a phantom scalar field possess
non-singular, spherically symmetry solutions, although a stability analysis on
these solutions indicates they are unstable.

The dark energy star models are also a generalization of a new
emerging picture for an alternative final state of gravitational
collapse, namely, the gravastar (\textit{grav}itational
\textit{va}cuum \textit{star}) models. The latter proposed by
Mazur and Mottola \cite{Mazur}, has an effective phase transition
at/near where the event horizon is expected to form, and the
interior is replaced by a de Sitter condensate. The latter is then
matched to a thick layer, with an equation of state given by
$p=\rho$, which is in turn matched to an exterior Schwarzschild
solution. The issue of gravastars has been extensively analyzed in
the literature, and we refer the reader to Refs.
\cite{gravastar2,ref:contgrava}. The generalization of the
gravastar picture is considered by matching an interior solution
governed by the dark energy equation of state, $\omega\equiv p/
\rho<-1/3$, to an exterior Schwarzschild vacuum solution at a
junction interface \cite{Lobo:2005uf}. The dynamical stability of
the transition layer was also explored, and it was found that
large stability regions exist that are sufficiently close to where
the event horizon is expected to form, so that it was argued that
it would be difficult to distinguish the exterior geometry of the
dark energy stars from an astrophysical black hole. Thus, these
alternative models do not possess a singularity at the origin and
have no event horizon, as its rigid surface is located at a radius
slightly greater than the Schwarzschild radius. This restriction
arises from the observed lack of energy emission due to surface
collisions of infalling material in suspected black hole systems.
In fact, although evidence for the existence of black holes is
very convincing, a certain amount of scepticism regarding the
physical reality of event horizons is still encountered, and it
has been argued that despite the fact that observational data do
indeed provide strong arguments in favor of event horizons, they
cannot fundamentally prove their existence \cite{AKL}.

As mentioned above, recent fits to observational data probably
favor an evolving equation of state, with the dark energy
parameter crossing the phantom divide $\omega=-1$ \cite{Vikman}.
Motivated by this fact, in a rather speculative scenario one may
theoretically consider the existence of a dark energy star, with
an evolving parameter starting out in the range $-1<\omega<-1/3$,
and crossing the phantom divide, $\omega=-1$. Once in the phantom
regime, the null energy condition is violated, which physically
implies that the negative radial pressure exceeds the energy
density. Therefore, an enormous negative pressure in the center
may, in principle, imply a topology change, consequently opening
up a tunnel, and converting the dark energy star into a wormhole
\cite{Lobo:2005uf,Morris}. One may assume that the topology change
may occur at approximately the Planck length scales, and once
created may be self-sustained as shown in Ref.
\cite{Garattini:2007ff}. In fact, the change in topology is an
extremely subtle issue, as in general relativity these changes
probably entail spacetime singularities. However, at the Planck
length scales quantum gravity effects dominate and spacetime
undergoes a deep and rapid transformation in its structure,
probably producing a multiply-connected quantum foam structure
\cite{Wheeler,geons}. It was suggested in Ref. \cite{Morris} that
one could imagine an absurdly advanced civilization
\cite{Lemos:2003jb} pulling a wormhole from this submicroscopic
spacetime quantum foam and enlarging it to macroscopic dimensions.
However, in a more plausible scenario, the possibility that
inflation might provide a natural mechanism for the enlargement of
such wormholes to macroscopic size was explored
\cite{Roman:1992xj}. In this work, we outline the theoretical
difficulties associated to the change in topology and present a
method based on the Casimir energy approach. Although it is still
unsure if this method produces a topology change, it is extremely
useful as the quasi-local energy difference calculation may
reflect or measure the occurrence of a change in topology. Other
concepts of topology changing spacetimes have been studied, for
instance: Using semi-classical and Morse-index methods in Refs.
\cite{Morseindex}; higher order back-reaction terms due to
fluctuations of gauge fields in the vicinity of a black hole may
result in the formation of a wormhole-like object
\cite{Hochberg:1992rd}; and more recently an approach based on a
Ricci flow may result in quantum wormholes
\cite{Dzhunushaliev:2008cz}.

Once the topology change has occurred, with the respective opening of a
tunnel, then the dark energy star has been converted into wormhole supported
by phantom energy. As mentioned above, it has recently been shown that
traversable wormholes may, in principle, be supported by phantom energy
\cite{Sushkov,Lobo-phantom}, which apart from being used as interstellar
shortcuts, may induce closed timelike curves with the associated causality
violations \cite{mty,Visser}. Particularly interesting solutions were found
\cite{Lobo-phantom}, and by using the ``volume integral quantifier'', it was
found that these wormhole geometries are, in principle, sustained by
arbitrarily small amounts of averaged null energy condition (ANEC) violating
phantom energy. A complementary approach was traced out in \cite{Sushkov}, by
considering specific choices for the distribution of the energy density
threading the wormhole. Recently, $4D$ static wormhole solutions supported by
two interacting phantom fields were found as well \cite{Dzhunushaliev:2007cs}.
Despite the fact that traversable wormholes violate the NEC in general
relativity (see Ref. \cite{Lobo:2007zb} for a recent review), it has been
shown that the stress energy tensor profile may satisfy the energy conditions
in the throat neighborhood in dynamic wormholes (see Ref.
\cite{Arellano:2006ex} and references therein) and in certain alternative
theories to general relativity \cite{satisfyNEC}. Perhaps not so appealing,
one could denote these exotic geometries consisting of dark energy stars (in
the phantom regime) and phantom wormholes as \textit{phantom stars}. We would
like to state our agnostic position relatively to the existence of dark energy
stars and phantom wormholes, or for that matter of \textit{phantom stars}.
However, it is important to understand their general properties and
characteristics, and we emphasize that the presence of a dark energy fluid
permeating the universe makes the study of dark energy condensates a
physically relevant endeavor.

This paper is organized in the following manner: In section \ref{secII}, we
briefly review static dark energy stars, followed by a deduction of general
solutions of time-dependent spacetimes. In Section \ref{secIII}, specific
time-dependent dark energy solutions are outlined, in particular, we present
the specific cases of a constant energy density, the Tolman-Matese-Whitman
mass function solution, and a class of models with a non-zero energy flux
term, which form from gravitational collapse. In Section \ref{secIV}, we
describe the theoretical difficulties associated with changes in topology and
present in some detail specific methods used in the literature, namely, a
Casimir energy approach involving quasi-local energy difference calculations
that may reflect or measure the occurrence of a topology change. We also briefly review the Morse index analysis. In section
\ref{sec:conclusion}, we conclude.


\section{Time-dependent dark energy stars}

\label{secII}


\subsection{Static spacetime}


In this section, we provide a brief outline of the mathematical models of
static and spherically symmetric dark energy stars considered in Ref.
\cite{Lobo:2005uf}. Consider the following time-independent line element, in
curvature coordinates, representing a dark energy star
\begin{equation}
ds^{2}=-e^{2\alpha(r)}\,dt^{2}+\frac{dr^{2}}{1-2m(r)/r}+r^{2}\,(d\theta
^{2}+\sin^{2}{\theta}\,d\phi^{2})\,, \label{ds metric}%
\end{equation}
where $\alpha(r)$ and $m(r)$ are arbitrary functions of the radial coordinate,
$r$. The function $m(r)$ can be interpreted as the quasi-local mass, and is
denoted as the mass function \cite{Lobo:2005uf}. The factor $\alpha(r)$ is the
\textquotedblleft gravity profile\textquotedblright\ and is related to the
locally measured acceleration due to gravity, through the following
relationship: $\mathcal{A}=\sqrt{1-2m(r)/r}\,\alpha^{\prime}(r)$
\cite{Lobo:2005uf,Lobo:2006ue}, where the prime denotes a derivative with
respect to the radial coordinate $r$. The convention used is that
$\alpha^{\prime}(r)$ is positive for an inwardly gravitational attraction, and
negative for an outward gravitational repulsion.

The Einstein field equations are given by \cite{Lobo:2005uf}
\begin{align}
m^{\prime}  &  =4\pi r^{2} \rho\,,\label{mass}\\
\alpha^{\prime}  &  =\frac{m+4\pi r^{3} p_{r}}{r(r-2m)} \,,\label{Phi}\\
p_{r}^{\prime}  &  =-\frac{(\rho+p_{r})(m+4\pi r^{3} p_{r})}{r(r-2m)}
+\frac{2}{r}(p_{t}-p_{r}) \,, \label{anisotTOV}%
\end{align}
where $\rho(r)$ is the energy density, $p_{r}(r)$ is the radial pressure, and
$p_{t}(r) $ is the tangential pressure orthogonal to $p_{r}$. Note that Eq.
(\ref{anisotTOV}) corresponds to the anisotropic pressure
Tolman-Oppenheimer-Volkoff (TOV) equation.

An additional constraint is placed on the system of equations by considering
the dark energy equation of state, $p_{r}(r)=\omega\rho(r)$, and taking into
account Eqs. (\ref{mass}) and (\ref{Phi}), we have the following relationship
\begin{equation}
\alpha^{\prime}(r)=\frac{m+\omega rm^{\prime}}{r\,\left(  r-2m \right)  } \,.
\label{EOScondition}%
\end{equation}
There is, however, a subtle point that needs to be emphasized
\cite{Sushkov,Lobo-phantom}. The notion of dark energy is that of a spatially
homogeneous cosmic fluid. Nevertheless, it can be extended to inhomogeneous
spherically symmetric spacetimes, by regarding that the pressure in the
equation of state $p=\omega\rho$ is a radial pressure, and that the transverse
pressure may be obtained from Eq. (\ref{anisotTOV}). In addition to this, and
as mentioned in the Introduction, despite the fact that the dark energy
equation of state represents a spatially homogeneous cosmic fluid and is
assumed not to cluster, inhomogeneities may arise due to gravitational
instabilities. Thus, the dark energy star geometries considered here may
possibly originate from density fluctuations in the cosmological background,
resulting in the nucleation through the respective density perturbations
\cite{Lobo:2006ue}.

In Ref. \cite{Lobo:2005uf}, specific solutions were found by considering that
the energy density is positive and finite at all points in the interior of the
dark energy star. In particular, several relativistic dark energy stellar
configurations were analyzed by imposing specific choices for the mass
function $m(r)$, and through Eq. (\ref{EOScondition}), $\alpha(r)$ was
determined, consequently providing explicit expressions for the stress-energy
tensor components. This interior solution was further matched to an exterior
Schwarzschild vacuum solution given by
\begin{equation}
ds^{2}=-\left(  1-\frac{2M}{r}\right)  \,dt^{2} +\left(  1-\frac{2M}
{r}\right)  ^{-1}dr^{2}+r^{2} \,(d\theta^{2}+\sin^{2}{\theta} \, d\phi^{2})
\,, \label{Sch-metric}%
\end{equation}
at a junction interface $a$. The Schwarzschild spacetime possesses an event
horizon at $r_{b}=2M$, so that to avoid the latter, the junction radius lies
outside $2M$, i.e., $a>2M$.

The surface stresses on the thin shell are given by
\begin{align}
\sigma &  =-\frac{1}{4\pi a} \left(  \sqrt{1-\frac{2M}{a}+\dot{a}^{2}}-
\sqrt{1-\frac{2m}{a}+\dot{a}^{2}} \, \right)  ,\label{surfenergy}\\
\mathcal{P}  &  =\frac{1}{8\pi a} \Bigg(\frac{1-\frac{M}{a} +\dot{a}
^{2}+a\ddot{a}}{\sqrt{1-\frac{2M}{a}+\dot{a}^{2}}} - \frac{1+\omega m^{\prime
}-\frac{m}{a}+\dot{a}^{2}+a\ddot{a}+\frac{\dot{a}^{2} m^{\prime}(1+\omega
)}{1-2m/a}} {\sqrt{1-\frac{2m}{a}+\dot{a}^{2}}} \, \Bigg) \,,
\label{surfpressure}%
\end{align}
where $\sigma$ and $\mathcal{P}$ are the surface energy density and the
tangential surface pressure \cite{Lobo:2005uf,WHshell,WHshell2}, respectively.
The overdot denotes a derivative with respect to $\tau$, which is the proper
time on the junction interface, and the prime here denotes a derivative with
respect to the junction surface radius $a$.

The dynamical stability of the transition layer $a$ of these dark energy stars
to linearized spherically symmetric radial perturbations about static
equilibrium solutions was also explored. It was found that large stability
regions exist that are sufficiently close to where the event horizon is
expected to form, so that it would be difficult to distinguish the exterior
geometry of the dark energy stars, analyzed in \cite{Lobo:2005uf}, from an
astrophysical black hole.


\subsection{Time-dependent spacetime}


In this section, we generalize the above static dark energy star models to
time-dependent geometries. This is mainly motivated by the fact that recent
fits to supernovae, CMB and weak gravitational lensing data probably favor an
evolving equation of state, with the dark energy parameter crossing the
phantom divide $\omega=-1$ \cite{Vikman}.

In the following, we consider a time-dependent and spherically symmetric
metric given by
\begin{equation}
ds^{2}=-e^{2\alpha(r,t)}dt^{2}+e^{2\beta(r,t)}dr^{2}+r^{2}\left(  d\theta
^{2}+\sin^{2}\theta d\phi^{2}\right)  \,. \label{metr1}%
\end{equation}
Note that one may also define the function $\beta(r,t)$ as
\begin{equation}
\beta(r,t)=\frac{1}{2}\ln\left[  1-\frac{2m(r,t)}{r}\right]  \,,
\end{equation}
where the `mass function' $m(r,t)$ is now time-dependent.

The Einstein field equation provides the following nonzero components:
\begin{align}
\rho(r,t)  &  =\frac{e^{-2\beta}}{8\pi r^{2}}(2\beta^{\prime}r+e^{2\beta
}-1)\,,\label{fieldeq1}\\
p_{r}(r,t)  &  =\frac{e^{-2\beta}}{8\pi r^{2}}(2\alpha^{\prime}r-e^{2\beta
}+1)\,,\label{fieldeq2}\\
f(r,t)  &  =\frac{\dot{\beta} e^{-(\alpha+\beta)}}{4\pi r}\,,
\label{fieldeq2b}\\
p_{t}(r,t)  &  =\frac{1}{8\pi}\Bigg\{e^{-2\beta}\left[  \alpha^{\prime\prime
}+\alpha^{\prime2}-\alpha^{\prime}\beta^{\prime}+\frac{1}{r}\left(
\alpha^{\prime}-\beta^{\prime}\right)  \right]  +e^{-2\alpha}\left(
\dot{\alpha}\dot{\beta}-\ddot{\beta}-\dot{\beta}^{2}\right)  \Bigg\} \,,
\label{fieldeq3}%
\end{align}
where the prime denotes a partial derivative with respect to the radial
coordinate $r$, and the overdot a partial derivative with respect to the time
coordinate $t$. Note the presence of an energy flux term in the radial
direction, $T^{t}{}_{r}=\pm f(r,t)$, which depends on $\dot{\beta}$.

An important issue in the time-dependent dark energy stars with the parameter
crossing the phantom divide are the energy conditions, in particular, the null
energy condition (NEC). The NEC is defined as $T_{\mu\nu}k^{\mu}k ^{\nu}\geq
0$, where $k^{\mu}$ is any null vector, and consequently provides $\rho
+p_{r}\pm2f \geq0$. The latter definition, taking into account the field
equations (\ref{fieldeq1})-(\ref{fieldeq2b}), is given by
\begin{equation}
\label{nec}\rho+p_{r}\pm2f=\frac{1}{4\pi r}\left[  e^{-2\beta}(\alpha^{\prime
}+\beta^{\prime})\pm2\dot{\beta}e^{-(\alpha+\beta)}\right]  \geq0 \,.
\end{equation}

In this context, one may consider a generalization of the equation of state
$p_{r}=\omega\rho$, given by
\begin{equation}
p_{r}(r,t)=\omega(r,t)\left[  \rho(r,t)\pm2f(r,t)\right]  \,, \label{eqstate1}%
\end{equation}
or
\begin{equation}
p_{r}(r,t)\mp2f(r,t)=\omega(r,t)\rho(r,t) \,, \label{eqstate2}%
\end{equation}
where $f(r,t)=\mp T^{t}{}_{r}$ is the energy flux term, as noted above.
However, one can come up with an interesting class of solutions considering
the following equation of state:
\begin{equation}
\label{darkEOS}p_{r}(r,t)=\omega(r,t)\rho(r,t) \,.
\end{equation}
Throughout this work, we essentially use the equation of state given by Eq.
(\ref{darkEOS}).

Taking into account the field equations (\ref{fieldeq1}) and (\ref{fieldeq2}),
then Eq. (\ref{darkEOS}) provides the following relationship
\begin{equation}
\label{timeEOS}\omega(r,t)\beta^{\prime}(r,t)=\alpha^{\prime}(r,t)+\frac
{1}{2r}[1+\omega(r,t)] \left[  1-e^{2\beta(r,t)}\right]  \,.
\end{equation}
A similar analysis was carried out in Ref. \cite{Kuhfittig}, in the context of
time-dependent wormholes.

Equation (\ref{timeEOS}) may be formally solved in terms of $\alpha(r,t)$, and
provides the following general solution
\begin{equation}
\label{gen:alpha}\alpha(r,t)=\int\frac{1}{\bar{r}}\left\{  \omega(\bar
{r},t)\beta^{\prime}(\bar{r},t)\bar{r} -\frac{1}{2} \left[  1+\omega(\bar
{r},t)\right]  \left[  1-e^{2\beta(\bar{r},t)} \right]  \right\}  \,d\bar{r}
\,.
\end{equation}
Thus, in principle, if $\omega(r,t)$ and $\beta(r,t)$ are known, then
$\alpha(r,t)$ may be obtained from Eq. (\ref{gen:alpha}).

As an alternative, Eq. (\ref{timeEOS}) can be formally integrated for
$\beta(r,t)$ to yield the general solution
\begin{align}
\label{timeEOSsol}\beta(r,t)  &  =-\frac{1}{2}\ln\left[  -2F(t) +\int
\frac{e^{\Gamma(\bar{r},t)}(1+\omega(\bar{r},t))} {\omega(\bar{r},t)\bar{r}
}\;d\bar{r} \right]  + \frac{1}{2}\Gamma(r,t) \,,
\end{align}
where $F(t)$ is an integration function, and the factor $\Gamma(r,t)$ is
defined as
\begin{equation}
\label{Gammasol}\Gamma(r,t)=\int\left[  \frac{2\bar{r}\alpha^{\prime}(\bar
{r},t)+\bar{r}+1}{\bar{r}\omega(\bar{r},t)} \right]  \;d\bar{r} \,.
\end{equation}

A particularly simple and interesting toy model is the specific case of a
purely time dependent parameter $\omega=\omega(t)$, so that the general
solution (\ref{timeEOSsol}) takes the form
\begin{align}
\label{timeEOSsol2}\beta(r,t)  &  =-\frac{1}{2}\Bigg\{\frac{2\alpha
(r,t)}{\omega(t)} -\ln\Big[r^{-\frac{1+\omega(t)}{\omega(t)}}\Big(-2F(t)
+\frac{1+\omega(t)}{\omega(t)}\int e^{\frac{2\alpha(\bar{r},t)}{\omega(t)}}
\bar{r}^{\frac{1}{\omega(t)}}\;d\bar{r}\Big) \Big]\Bigg\} \,.
\end{align}

In the next section, we analyze specific solutions, namely, that of a constant
energy density, the Tolman-Matese-Whitman mass function, which were
extensively explored in Ref. \cite{Lobo:2005uf}, and a collapsing model with a
nonzero energy flux term.


\section{Specific time-dependent dark energy star solutions}

\label{secIII}


\subsection{Constant energy density}


Consider the specific case of a constant energy density, $\rho(r,t)=\rho_{0}$,
so that Eq. (\ref{fieldeq1}) provides the solution
\begin{equation}
\beta(r,t)=-\frac{1}{2}\ln\left[  1-Ar^{2}+\frac{F_{1}(t)}{r}\right]  \,,
\end{equation}
with $A=8\pi\rho_{0}/3$ and $F_{1}(t)$ is a function of integration.

Substituting into Eq. (\ref{gen:alpha}), one arrives at
\begin{equation}
\label{gen2:alpha}\alpha(r,t) = \int\frac{3A\omega(r,t) r^{3}+Ar^{3}
-3F_{1}(t)}{2r\left[  3r-Ar^{3}+3F_{1}(t)\right]  }+F_{2}(t) \,,
\end{equation}
where $F_{2}(t)$ is another function of integration. One may further simplify
the analysis by considering that $F_{1}(t)=0$, which is physically justified
by the imposition of a finite mass function at the origin $r=0$ for all values
of the time coordinate $t$. Note that considering $F_{1}(t)=0$ implies that
the mass function is not time-dependent, and the flux term $f(r,t)$ is zero,
as $\dot{\beta}=0$.

For instance, consider the specific example of a separation of variables of
the parameter $\omega(r,t)$ given by
\begin{equation}
\omega(r,t)=\omega_{1}(t)\,\omega_{2}(r)\,. \label{om}%
\end{equation}
Choosing the following functions:
\begin{align}
\omega_{1}(t)  &  =\omega_{0}+\bar{\omega}_{0}\tanh\left[  \sigma
(t-t_{0})\right]  \,,\label{om1}\\
\omega_{2}(r)  &  =-\frac{\lambda^{2}}{1+(r/R)^{2}} \,, \label{om2}%
\end{align}
where $\omega_{0}$, $\bar{\omega}_{0}$, $\sigma$, $t_{0}$, $\lambda^{2}$, and
$R$ are constants. The factor $1/\sigma$ may be interpreted as the
``relaxation time'', describing the width of the time-dependence.

\begin{figure}[h]
\centering
\includegraphics[width=2.7in]{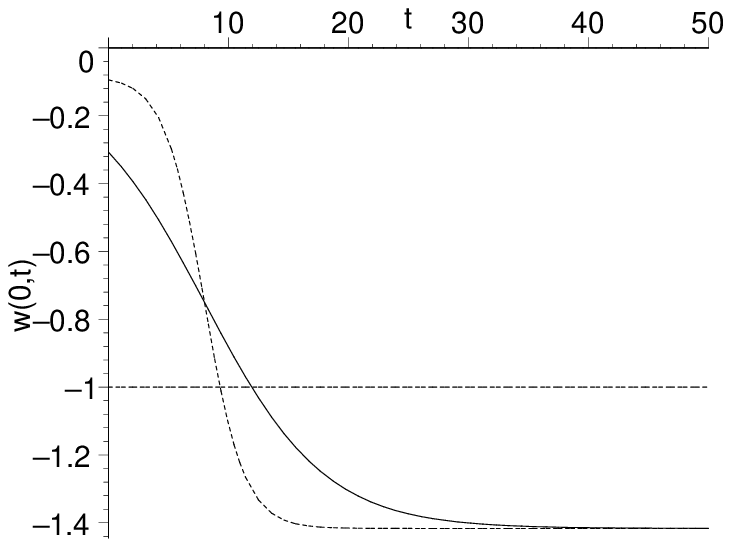} \hspace{0.4in}
\includegraphics[width=2.8in]{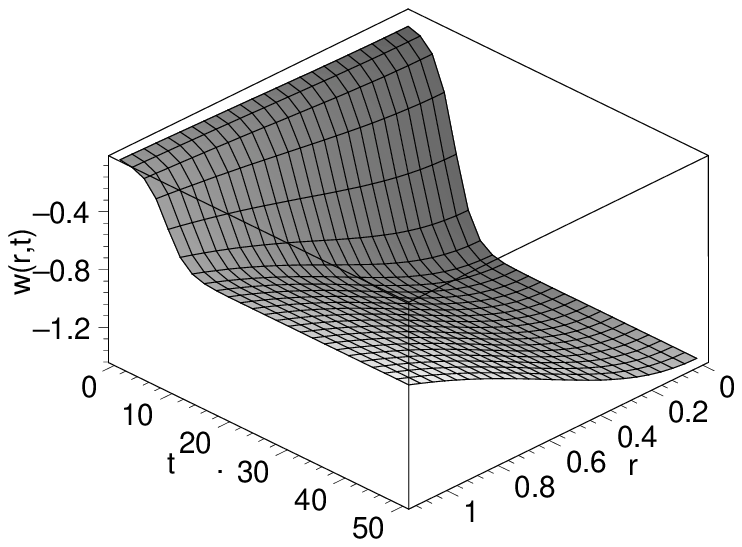}\caption{Left plot: For this case we
considered the behavior of $\omega=\omega(0,t)$ at the center $r=0$; for the
dashed curve we have $\sigma=0.3$ and for the solid curve $\sigma=0.1$. Right
plot: For this case we considered the $\omega=\omega(t,r)$ dependence with
$\sigma=0.3$. For both cases, we have assumed the following numerical values:
$R=1$, $\delta=1$, $\omega_{0}=3/4$, $\bar{\omega}_{0}=2/3$ and $t_{0}=8$.}%
\label{Fig:omega}%
\end{figure}

See Fig. \ref{Fig:omega} for a qualitative description of $\omega(r,t)$ given
by Eqs. (\ref{om})-(\ref{om2}). The left plot represents the behavior of
$\omega=\omega(0,t)$ at the center $r=0$. We have considered the following
values: $R=1$, $\delta=1$, $\omega_{0}=3/4$, $\bar{\omega}_{0}=2/3$ and
$t_{0}=8$. 

Substituting the functions (\ref{om})-(\ref{om2}) into Eq. (\ref{gen2:alpha}),
yields the following solution
\begin{equation}
\alpha(r,t) = \frac{1}{4}\ln\left\{  \frac{(1-Ar^{2})^{\bar{A}\omega_{1}
(t)-1}} {\left[  1+(r/R)^{2}\right]  ^{\bar{A}\omega_{1}(t)}} \right\}
+F_{2}(t) \,,
\end{equation}
where, for notational simplicity, the constant $\bar{A}$ is defined as
\begin{equation}
\bar{A}=\frac{3AR^{2}\lambda^{2}}{R^{2}A+1} \,.
\end{equation}
The function $F_{2}(t)$ can be absorbed through a redefinition of the time
coordinate as before, so that without a significant loss of generality one may
impose the condition $F_{2}(t)=0$.

The pressure profile is given by
\begin{align}
p_{r}  &  =-\frac{3\lambda^{2}R^{2}A\omega_{1}(t)}{8\pi(R^{2}+r^{2})} \,,\\
p_{t}  &  =\frac{3A}{32\pi}\,\frac{R^{2}Ar^{2}(R^{2}+2r^{2})+Ar^{6}
+\omega_{1}(t)\lambda^{2}R^{2}[3Ar^{2}\omega_{1}(t)\lambda^{2}R^{2}
-4(Ar^{4}+R^{2})]} {(1-Ar^{2})(R^{2}+r^{2})^{2}} \,,
\end{align}
with $p_{r}=p_{t}$ at the center, $r=0$.

The analysis simplifies by considering a purely time-dependent parameter,
i.e., $\omega=\omega(t)$. Thus, Eq. (\ref{gen2:alpha}) takes the form
\begin{equation}
\alpha(r,t)=-\frac{1}{4}\left[  1+3\omega(t)\right]  \ln(1-Ar^{2})+F_{2}(t)
\,.
\end{equation}
The factor $F_{2}(t)$ can be absorbed into a redefinition of the time
coordinate, so that without a significant loss of generality, one can assume
$F_{2}(t)=0$.

The pressure profile is given by the following relationships:
\begin{align}
\label{pressprofile}p_{r}  &  =\frac{3A\omega(t)}{8\pi} \,,\nonumber\\
p_{t}  &  =\frac{3A\left[  4\omega(t)+Ar^{2}+3Ar^{2}\omega^{2}(t)\right]  }
{32\pi(1-Ar^{2})} \,.
\end{align}
Note that $p_{r}=p_{t}$ at the center, $r=0$, as expected.


\subsection{Tolman-Matese-Whitman mass function}


An interesting example is the Tolman-Matese-Whitman mass function considered
in Ref. \cite{Lobo:2005uf}. As in the example outlined above, we impose that
$\dot{\beta}=0$, so that the flux term $f(r,t)$ is zero. Thus, consider the
following choice for the time-independent mass function, given by
\begin{equation}
\beta(r,t)=\beta(r)=\frac{1}{2}\ln\left(  \frac{1+2b_{0}r^{2}}{1+b_{0}r^{2}
}\right)  \,, \label{TMWmass}%
\end{equation}
where $b_{0}$ is a non-negative constant \cite{Lobo:2005uf}. The latter may be
determined from the regularity conditions and the finite character of the
energy density at the origin $r=0$, and is given by $b_{0}=8\pi\rho_{c}/3$,
where $\rho_{c}$ is the energy density at $r=0$.

Now, consider the radial and temporal dependent case of $\omega=\omega(r,t)$
given by the functions (\ref{om})-(\ref{om2}). Substituting these functions
and Eq. (\ref{TMWmass}) into Eq. (\ref{gen:alpha}), yields the following
solution
\begin{equation}
\alpha(r,t) = \frac{1}{2}\ln\left\{  \left(  1+b_{0}r^{2}\right)  ^{\Sigma(t)}
\left(  1+2b_{0}r^{2}\right)  ^{\Upsilon(t)} \left[  1+\left(  \frac{r}{
R}\right)  ^{2}\right]  ^{\Xi(t)} \right\}  +F_{2}(t) \,,
\end{equation}
where the $F_{2}(t)$ is a function of integration which may be reabsorbed in a
redefinition of the time coordinate, so that without a loss of generality we
impose $F_{2}(t)=0$, as before. For notational simplicity, we have considered
the following definitions
\begin{align}
\Sigma(t)  &  =\frac{1+b_{0}R^{2}(2b_{0}R^{2}-3) -b_{0}\lambda^{2}R^{2}
\omega_{1}(t)(1-2b_{0}R^{2})} {2(1-2R^{2}b_{0})(1-R^{2}b_{0})} \,,\\
\Upsilon(t)  &  =\frac{2b_{0}\lambda^{2}R^{2}} {1-2R^{2}b_{0}} \,,\\
\Xi(t)  &  =\frac{b_{0}\lambda^{2}R^{2}\omega_{1}(t)(2b_{0}R^{2}-3)}
{2(1-2R^{2}b_{0})(1-R^{2}b_{0})} \,,
\end{align}
respectively.

The stress-energy tensor components are given by
\begin{align}
\rho(r)  &  =\frac{b_{0}(3+2b_{0}r^{2})}{8\pi(1+2b_{0}r^{2})^{2}} \,,\\
p_{r}(r,t)  &  =-\frac{b_{0}R^{2}\lambda^{2}(3+2b_{0}r^{2})\omega_{1}(t)}
{8\pi(R^{2}+r^{2})(1+2b_{0}r^{2})^{2}} \,,\\
p_{t}(r,t)  &  =b_{0} \Big[4R^{4}b_{0}^{3}r^{6}(1-\omega_{1}(t)\lambda
^{2})^{2} +8R^{2}b_{0}^{3}r^{8}(1+\omega_{1}(t)\lambda^{2}) +4R^{4}b_{0}
^{2}r^{4}(2-3\omega_{1}(t)\lambda^{2}+3\omega_{1}(t)\lambda^{4})\nonumber\\
&  +4R^{2}b_{0}^{2}r^{6}(4+9\omega_{1}(t)\lambda^{2}) +R^{4}b_{0}
r^{2}(3-16\omega_{1}(t)\lambda^{2}+9\omega^{2}_{1}(t)\lambda^{4}) +2R^{2}
b_{0}r^{4}(3+14\omega_{1}(t)\lambda^{2})\nonumber\\
&  +b_{0}r^{6}(3+8b_{0}r^{2}+4b_{0}^{2}r^{4})-12\omega_{1}(t)R^{4}\lambda^{2}
\Big]\Big/\Big[32\pi(1+b_{0}r^{2})(1+2b_{0}r^{2})^{3}(R^{2}+r^{2}
)^{2}\Big] \,.
\end{align}
Note that $p_{r}=p_{t}$ at the center, $r=0$.

For simplicity, considering a purely time-dependent parameter $\omega
=\omega(t)$, and substituting (\ref{TMWmass}) into Eq. (\ref{gen:alpha}),
provides the following solution
\begin{align}
\alpha(r,t)=\frac{1}{2}\ln\left[  \left(  1+b_{0}r^{2}\right)  ^{\frac
{1-\omega(t)}{2}} \left(  1+2b_{0}r^{2}\right)  ^{\omega(t)}\right]  +F(t) \,,
\end{align}
where the $F(t)$ is a function of integration which, as before, may be
absorbed into a redefinition of the time coordinate, so that one may consider
$F(t)=0$ without a significant loss of generality.

The stress-energy tensor components are given by
\begin{align}
p_{r}  &  =\omega(t)\rho=\frac{b_{0}(3+2b_{0}r^{2})\omega(t)}{8\pi
(1+2b_{0}r^{2})^{2}} \,,\\
p_{t}  &  =\frac{b_{0}\{4b_{0}^{2}r^{4}(1+\omega)[3+b_{0}r^{2}(1+\omega)]
+b_{0}r^{2}[3+\omega(9\omega+16)]+8b_{0}^{2}r^{4}+12\omega\}} {32\pi
(1+b_{0}r^{2})(1+2b_{0}r^{2})^{3}} \,,
\end{align}
with $p_{r}=p_{t}$ at the center, $r=0$.


\subsection{A class of models with a non-zero flux term}

\label{sec:IIIC}


In this section, we construct a set of models with a non-zero energy flux
term, where at early times possesses a small inhomogeneity in the region near
$r=0$ which grows due to gravitational collapse. For the specific case
considered in this section, we assume for simplicity that the parameter, which
eventually crosses the phantom divide in the central region, is purely
time-dependent, i.e., $\omega=\omega(t)$, and is governed by an equation of
state of the form
\begin{equation}
p_{r}(r,\,t)=\omega(t)\rho(r,\,t)\,, \label{eq:fluxeos}%
\end{equation}
and that the system tends to isotropy for large $r$.

For the energy density, we generalize the Mbonye-Kazanas density profile
\cite{ref:MB} (also utilized by Dymnikova \cite{ref:dym}) to a reasonable
time-dependent model given by
\begin{equation}
\rho(r,\,t)=\rho_{0}a(t)e^{-(r/r_{0})^{n}} \,, \label{MBrho}%
\end{equation}
where $\rho_{0}$, $r_{0}$ and $n$ are appropriately chosen constants. Here,
the time-dependent function $a(t)$ is chosen so that the collapse will
asymptote at late times, forming a static star. Note for the sake of clarity
that the time-dependent function $a(t)$ should not be confused with the
junction interface radius introduced in Eqs. (\ref{surfenergy})-(\ref{surfpressure}). This profile has been extremely useful in the
investigations of non-singular black holes (i.e., horizons not shielding a
singularity) \cite{ref:MB}, including de Sitter core black holes
\cite{ref:dym} and, more recently, as a model for gravastars
\cite{ref:contgrava} (supplemented with an appropriate equation of state).

The equation of state (\ref{eq:fluxeos}) then yields
\begin{equation}
p_{r}(r,\,t)=\omega(t)\rho_{0}a(t)e^{-(r/r_{0})^{n}}\, , \label{tdeppr}%
\end{equation}
At this stage, it is useful to write the solution to the field equations as
follows \cite{ref:ddt}:
\begin{subequations}
\begin{align}
e^{-2\beta(r,t)}=  &  \;1-\frac{8\pi}{r} \left[  b^{2}(t)+\int_{0_{+}}^{r}
\rho(\bar{r},t)\,\bar{r}^{2}\,d\bar{r}\right]  \,,\label{eq:invgrr}\\
e^{2\alpha(r,t)}=  &  \;e^{-2\beta(r,t)} \left\{  \exp\left[  h(t)+8\pi
\int_{0_{+}}^{r} \left[  p_{r}(\bar{r},t)+\rho(\bar{r},t)\right]
e^{2\beta(\bar{r},t)} \bar{r}\,d\bar{r}\right]  \right\}  \,,\label{eq:gtt}\\
f(r,t)=  &  \;-\frac{1}{r^{2}} \left[  2b(t)\dot{b}(t) +\int_{0_{+}}^{r}%
\dot{\rho}(\bar{r},t)\, \bar{r}^{2}\,d\bar{r}\right]  e^{2\left[  \beta(r,\,t)
-\alpha(r,\,t)\right]  } \,,\label{eq:energyflux}\\
p_{t}(r,t)=  &  \;\frac{r}{2}\left(  p_{r}^{\prime}+\dot{f}\right)  +\left(
1+\frac{r}{2}\alpha^{\prime}\right)  p_{r} +\frac{r}{2}\left(  \dot{\alpha}
+\dot{\beta}\right)  f +\frac{r}{2}\alpha^{\prime}\rho\;, \label{eq:pressure}%
\end{align}
respectively, where the explicit coordinate dependence has been dropped in Eq.
(\ref{eq:pressure}), and as before, the prime denotes a partial derivative
with respect to the radial coordinate $r$, and the overdot a partial
derivative with respect to the time coordinate $t$. The functions $b(t)$ and
$h(t)$ are two arbitrary functions of integration. For a star, $b(t)$ is set
to zero to avoid a singularity at the center. However, at late time this
function need not vanish as it is useful for the wormhole configuration. Note
that with the prescription of the energy density and radial pressure, the
entire system of equations may, in principle, be solved for the unknowns. We
exploit this fact here.

To ensure that the star crosses the phantom divide and yet does not collapse
for infinite time, the following prescriptions are made:
\end{subequations}
\begin{equation}
a(t)= a_{0}\left[  (1+\epsilon)-e^{-k_{0}t}\right]  ,\qquad\omega
(t)=\omega_{0}-\omega_{1}\left(  1-e^{-k_{1}t}\right)  , \label{eq:aandw}%
\end{equation}
where $0$ and $1$ subscripts denote constant quantities. We now
have an infinite family of solutions with the desired physical
properties. At this stage we should note that the generated
space-times tend to Minkowski space-time as $r\rightarrow\infty$.
One would need to therefore cut-off the solution at some $r=r_{*}$
and patch it to an appropriate dark energy exterior. However,
given how small the energy density and pressures of the currently
accelerating universe are (when compared to those of an average
star), the asymptotic Minkowski approximation is probably a
reasonable approximation.

\begin{figure}[h]
\centering
\includegraphics[height=2.3in, width=3.05in]{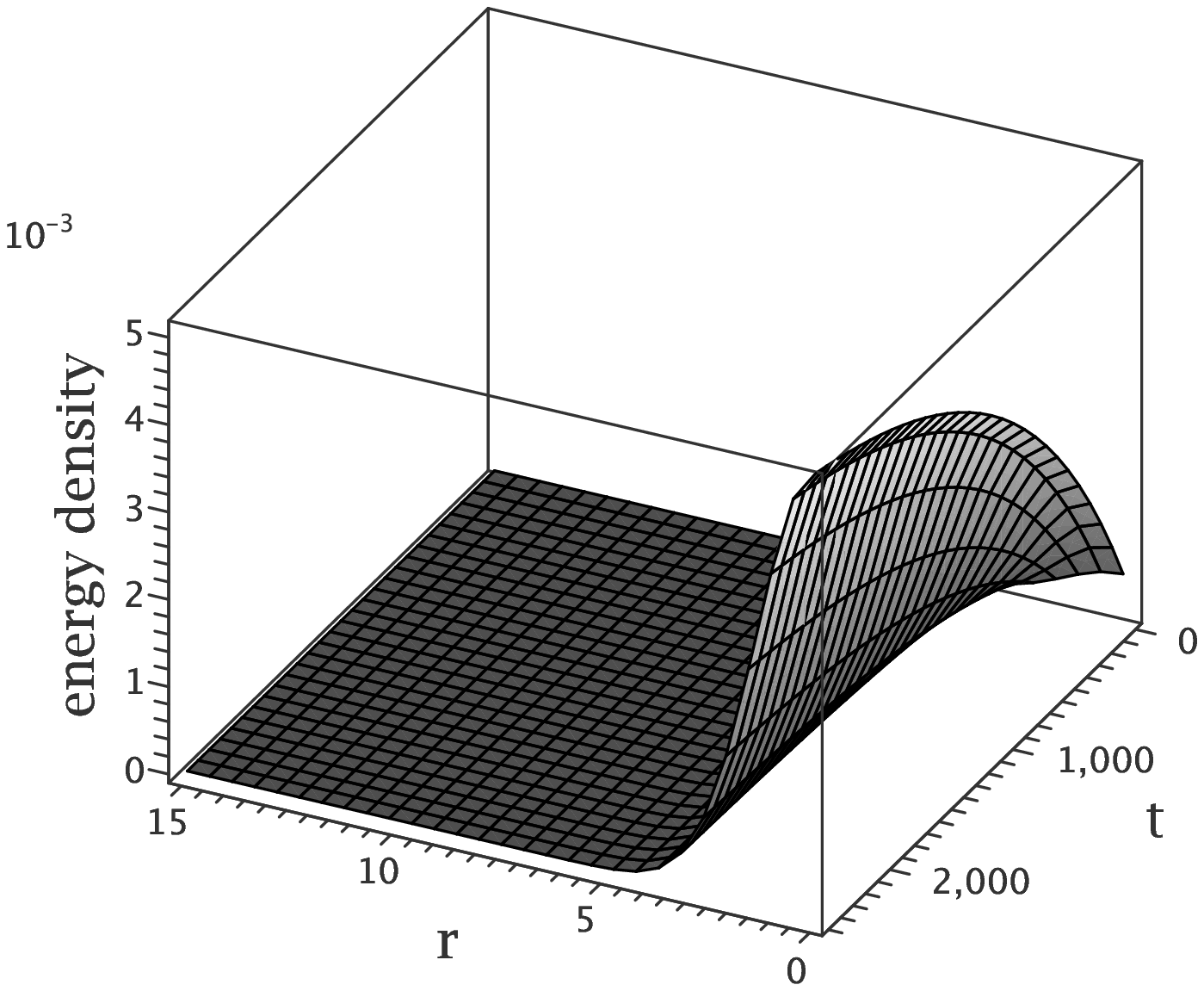} \vspace{-0.2in}
\includegraphics[height=2.3in, width=3.25in]{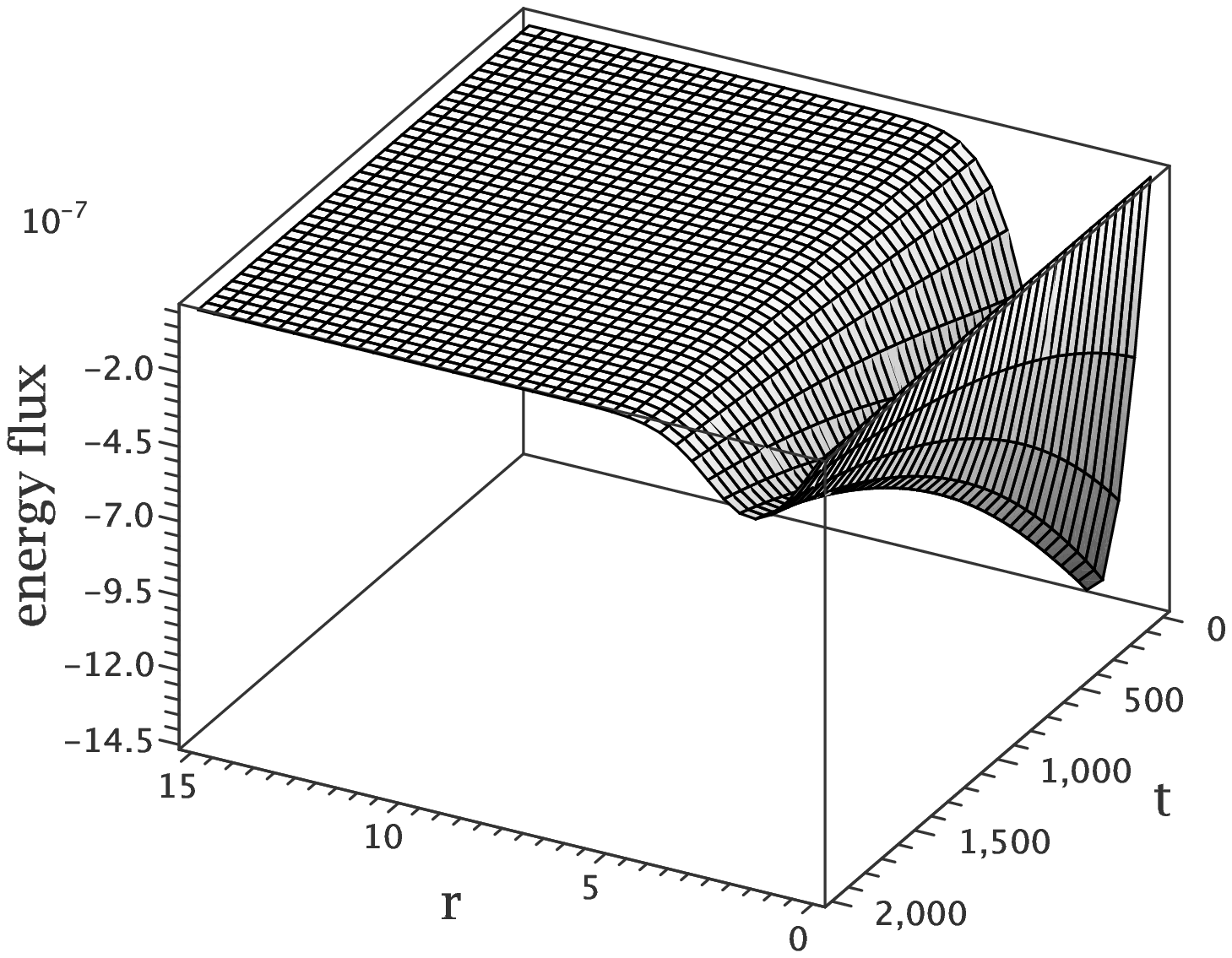}
\includegraphics[height=2.8in, width=3.25in]{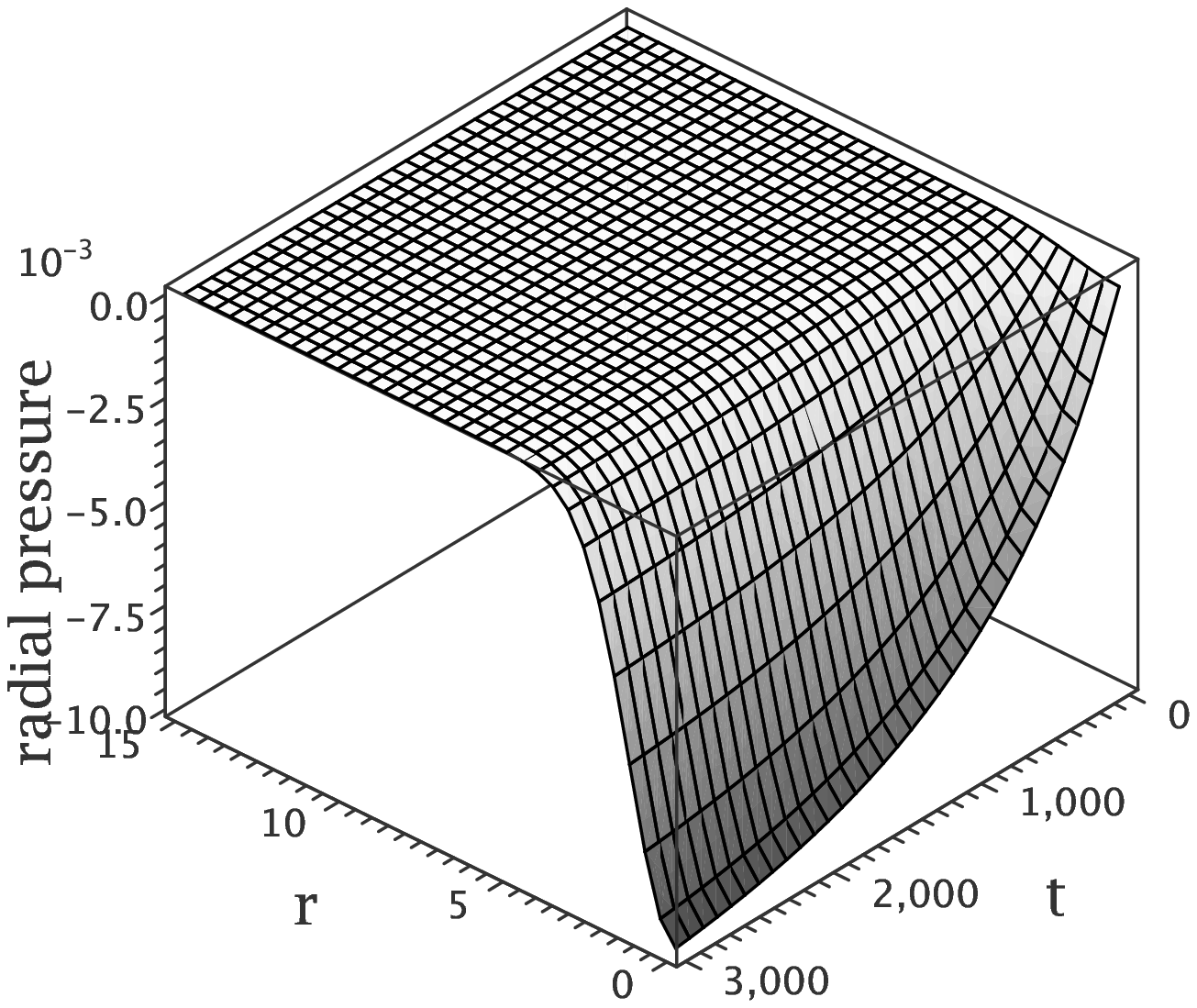} \hspace{-0.6in}
\vspace{-0.5in} \includegraphics[height=2.8in, width=4.0in]{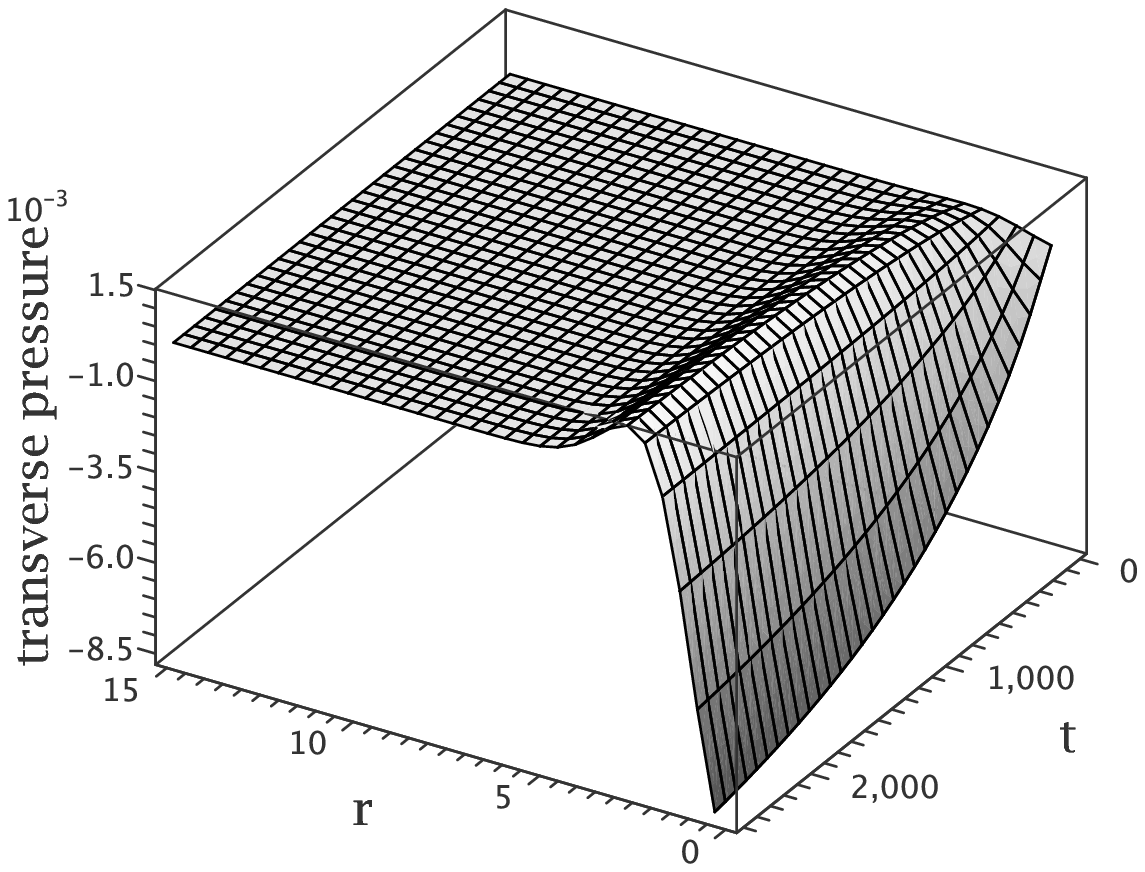}
\includegraphics[height=2.3in, width=2.8in]{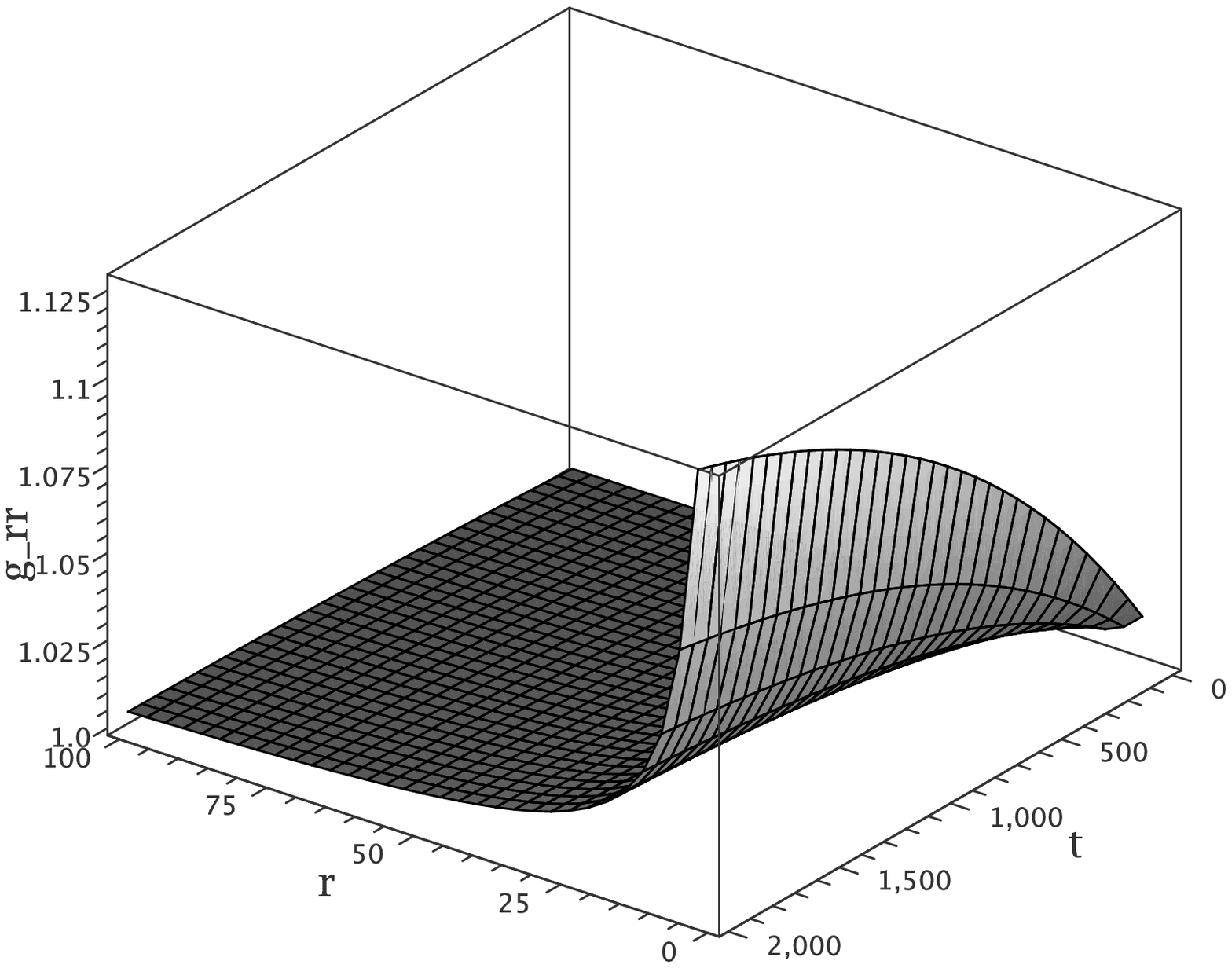} \hspace{0.3in}
\includegraphics[height=2.3in, width=2.87in]{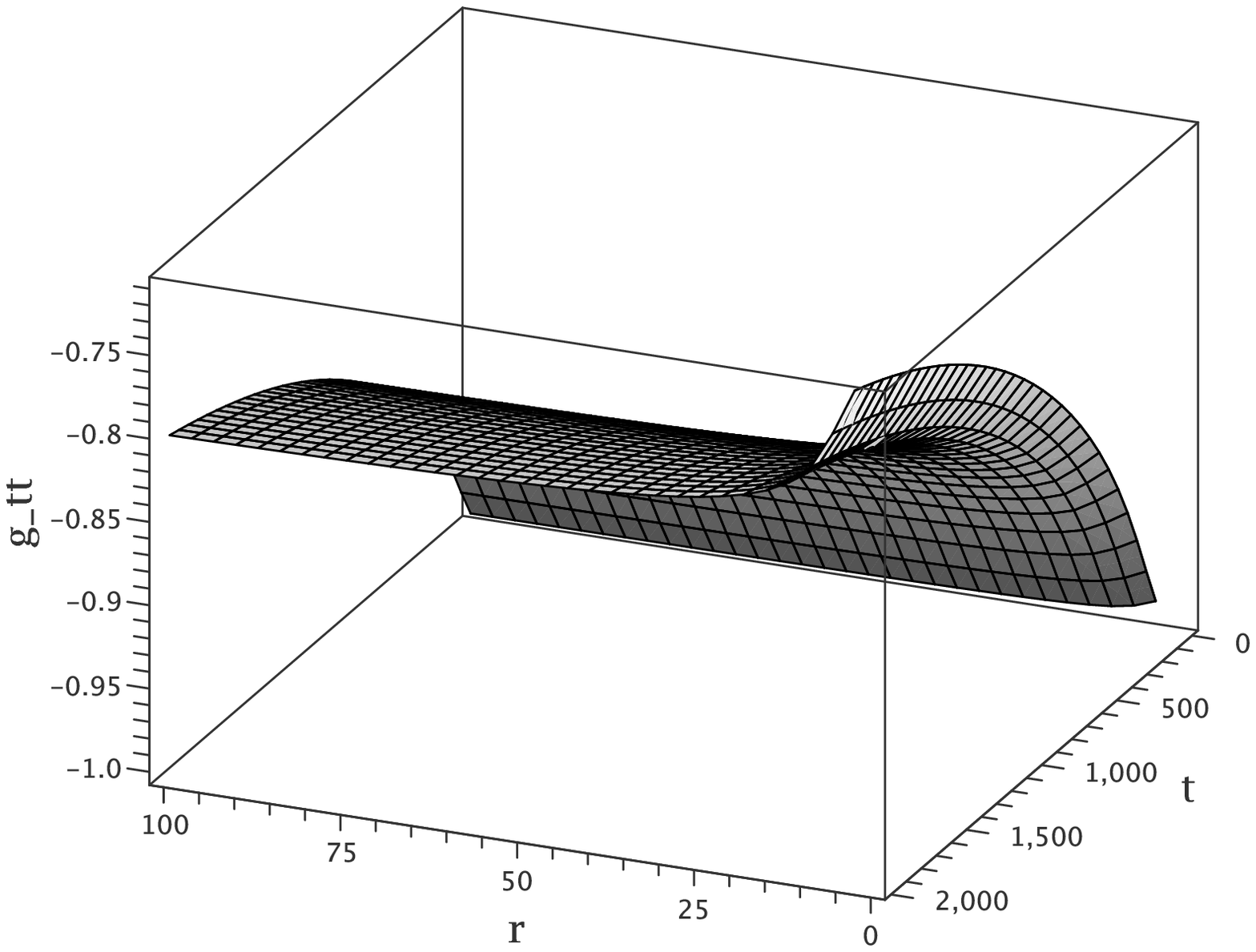}\caption{Parameters for
a sample stellar model presented in section III-C. Plotted on the vertical
axes (from left to right, starting from the top) are $\rho(r,\,t)$,
$T^{t}_{\;r}(r,\,t)$ $p_{r}(r,\,t)$, $p_{t}(r,\,t)$, and finally
$e^{2\beta(r,\,t)}$, and $-e^{2\alpha(r,\,t)}\,$.}%
\label{Fig:rho-pr}%
\end{figure}

\begin{figure}[h]
\centering
\includegraphics[height=1.85in, width=2.4in]{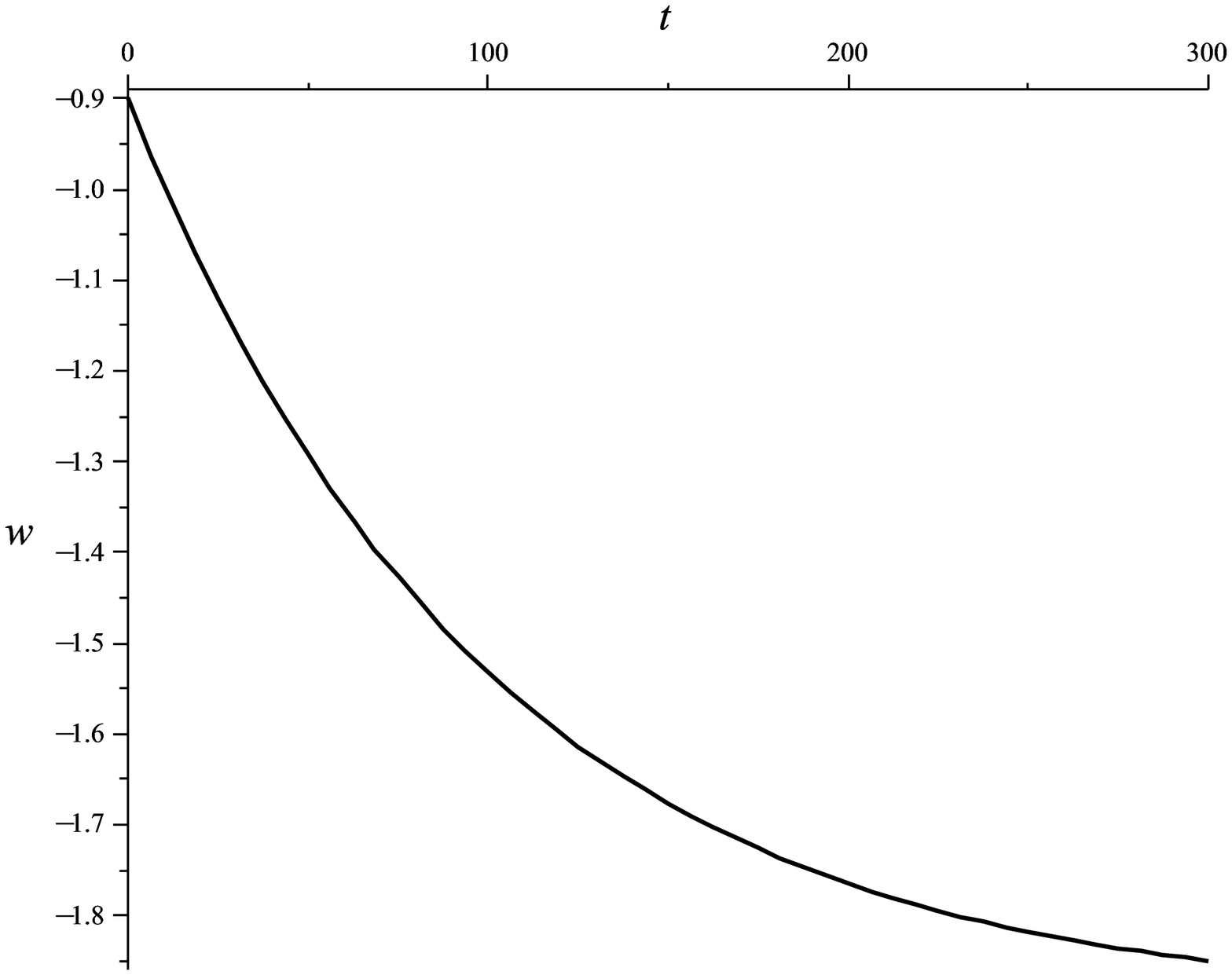} \hspace{0.6in}
\includegraphics[height=1.8in, width=2.4in]{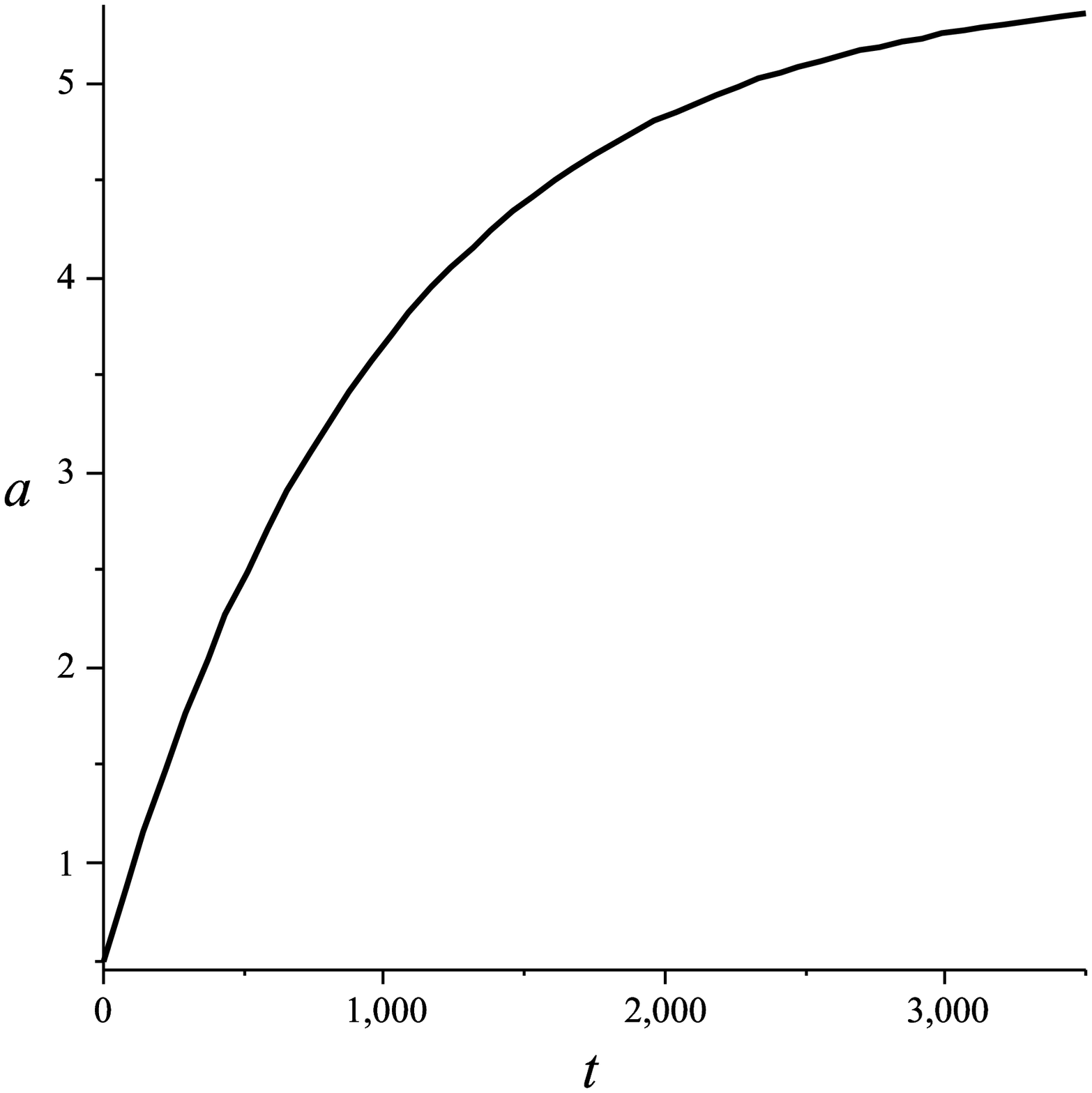}\caption{The parameters
$\omega(t)$ (left) and $a(t)$ (right) for a sample stellar model presented in
section III-C. For this particular evolution the following values were used:
$\omega_{0}=-0.9$, $k_{1}=0.01$, $a_{0}=5$, $\epsilon=0.1$, $k_{0}=0.001$,
$\omega_{1}=1$\,.}%
\label{Fig:wt}%
\end{figure}

\begin{figure}[h]
\centering
\includegraphics[width=3.0in]{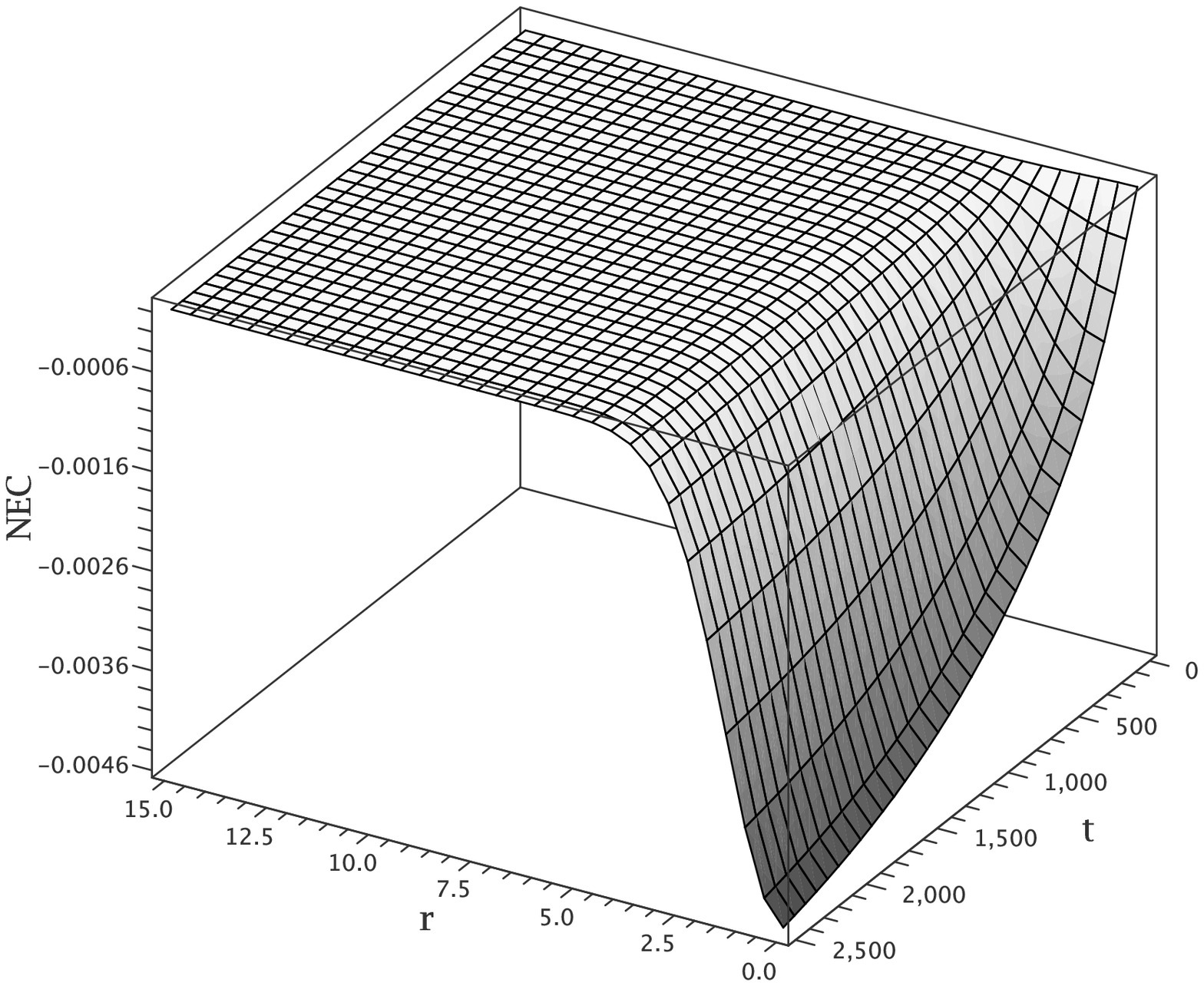}\caption{The figure depicts the
null energy condition, $\mathrm{NEC}=\rho+p_{r}+2f$ for the specific model
with a non-zero flux term considered in the text.}%
\label{fig:econd}%
\end{figure}

Surprisingly, for certain values of $n$, the equations (\ref{eq:invgrr}%
)-(\ref{eq:pressure}) may actually be integrated to yield an analytic result.
The expressions are rather unwieldly however so instead we plot the various
relevant parameters in figures \ref{Fig:rho-pr} and \ref{Fig:wt}. From the
figures it can be noted that the initial inhomogeneity is very small, most
pronounced near the center, and the space-time almost flat everywhere. There
is an inward flow of energy, due to the gravitational collapse as can be seen
by the negative values of the energy flux plot. At late time, the magnitude of
the energy flux decreases and asymptotes to zero, indicating the halt of the
collapse. At some point during the collapse, the phantom divide is crossed and
the conditions for possible wormhole formation are established. We show this
in figure \ref{fig:econd} where it may be seen that the NEC violation is most
severe in the center. 




\section{Topology change}

\label{secIV}

\subsection{The Casimir energy approach}

Although (\ref{eq:aandw}) also allows for an \emph{ordinary} (positive
pressure) fluid at early times, all of the geometries considered in the
previous sections were modelled so that the evolving parameter $\omega
(r,t)|_{r=0}$ starts out in the range $-1<\omega<-1/3$, then crosses the
phantom divide, and finally ends up in the phantom regime, $\omega=p_{r}%
/\rho<-1$. Once in the phantom regime, the negative radial pressure exceeds
the energy density, which in principle may imply a topology change. It is
still uncertain how to obtain this topology change, and if possible, is
riddled which difficulties, such as the theoretical appearance of closed
timelike curves. It is likely that crossing the phantom divide is accompanied
by a large quantum fluctuation of the metric. Then a crucial question is
\cite{Visser}: what happens when the metric fluctuations become large?

Concerning the geometry of spacetime undergoing quantum fluctuations, this
does not seem to be a source of disagreement, but when we turn to the question
of whether or not the topology of spacetime undergoes quantum fluctuations,
the problem becomes more subtle. It was J. A. Wheeler \cite{Wheeler,geons} who
first conjectured that spacetime could be subjected to a topology fluctuation
at the Planck scale. This means that spacetime undergoes a deep and rapid
transformation in its structure. The changing spacetime is best known as
spacetime foam, which can be taken as a model for the quantum gravitational
vacuum. Some authors have investigated the effects of such a foamy space on
the cosmological constant, for instance, one example is the celebrated Coleman
mechanism, where wormhole contributions suppress the cosmological constant,
explaining its small observed value \cite{Coleman}. Nevertheless, how to
realize such a foam-like space and also whether this represents the real
quantum gravitational vacuum are still unknown. We can mention some results
about topological constraints on the \textit{classical} evolution of general
relativistic spacetimes. They are summarized in two points \cite{Visser}:

\begin{enumerate}
\item In causally well-behaved classical spacetimes the topology of space does
not change as a function of time.

\item In causally ill-behaved classical spacetimes the topology of space can
sometimes change.
\end{enumerate}

From the \textit{quantum} point of view we can separate the problem of
topology change generated by a canonical quantization approach and a
functional integral quantization approach. The Hawking topology change theorem
is thus enough to show that the topology of space cannot change in canonically
quantized gravity \cite{Hawking:1991nk}. In the Feynman functional integral
quantization of gravitation things are different. Indeed, in this formalism,
an approach is possible to spacetime foam where we know that fluctuations of
topology become an important phenomenon at least at the Planck scale
\cite{Hawking:1978jz}. However, in our case we can adopt another strategy. In
some cases, we can create a one to one correspondence between topology and the
asymptotic energy. In particular, we will consider the Arnowitt-Deser-Misner
(ADM) energy \cite{ADM} as a reference energy. The reason for such a choice is
that $E_{ADM}\geq0$ and it is vanishing for flat space. Therefore we can think
about flat space as the unique reference space to compare a change in
spacetime associated to the corresponding topology. A trivial example could be
the comparison between flat space, where the topology is $R^{4}$ and the
Schwarzschild space, with topology $R^{2}\times S^{2}$: they are topologically
distinct and possess a distinct ADM energy: $E_{flat}=0$ and
$E_{Schwarzschild}=M$ . A topology transition from Schwarzschild to flat or
viceversa, should be necessarily accompanied by a change in ADM energy. In the
same manner, we can think that a transition from the dark star to the wormhole
could be associated to a change in the asymptotic energy, measured by the ADM
energy, namely if a topology change appears this could be reflected to a
change in the ADM energy. The way to detect this is simply computed by%
\begin{equation}
E_{ADM}^{DS}-E_{ADM}^{Wormhole}=\left(  E_{ADM}^{DS}-E_{ADM}^{Flat}\right)
-\left(  E_{ADM}^{Wormhole}-E_{ADM}^{Flat}\right)  \gtreqless0.
\label{ADM:DS-W}%
\end{equation}

\begin{figure}[h]
\centering
\includegraphics
[height=2.2085in, width=3.1713in] {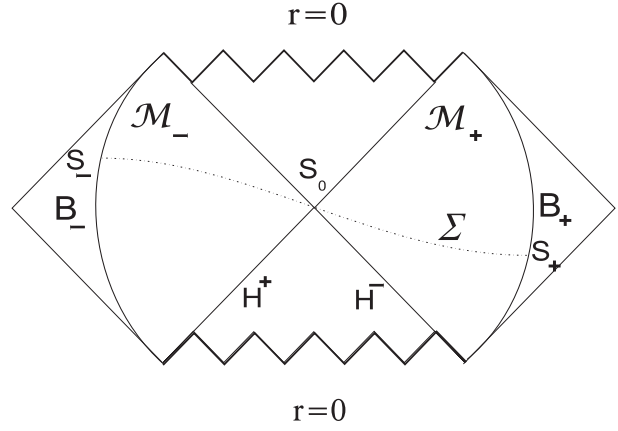}\caption{A
Penrose-Carter diagram illustrating the boundary location in a
Schwarzschild metric. $\mathcal{M}_{+}$ and $\mathcal{M}_{-}$ are
the two wedges, located in the right and left sectors of a Kruskal
diagram, respectively and bounded by two
boundaries $B_{+}$ and $B_{-}$ located in $\mathcal{M}_{+}$ and $\mathcal{M}%
_{-}$ respectively. $\Sigma=\Sigma_{+}\cup\Sigma_{-}$ is a spacelike
hypersurface. $H_{+}$ and $H_{-}$ are the future and past horizon,
respectively. $S_{0}$ $\left(  S_{0}=H_{+}\cap H_{-}\right)  $ is the
bifurcation surface (wormhole throat) and $S_{+}$ and $S_{-}$ are the
two-dimensional boundaries of $\Sigma_{+}$ and $\Sigma_{-}$, respectively.}%
\label{Penrosediagram}%
\end{figure}

For asymptotically flat spacetimes, the ADM energy is defined as%
\begin{equation}
E_{ADM}=\frac{1}{16\pi G}\int_{S}\left(  D^{i}h_{ij}-D_{j}h\right)
r^{j},\label{ADM}%
\end{equation}
where the indices $i,j$ run over the three spatial dimensions and%
\begin{equation}
h_{ij}=g_{ij}-\bar{g}_{ij},
\end{equation}
where $\bar{g}_{ij}$ is the background three-metric. $D_{j}\ $is the
background covariant derivative and $r^{j}$ is the unit normal to the large
sphere $S$. However, Hawking and Horowitz \cite{HawHor} have shown that the
definition $\left(  \ref{ADM}\right)  $ is equivalent to
\begin{equation}
E_{ADM}=\frac{1}{8\pi G}\int_{S^{\infty}}d^{2}x\sqrt{\sigma}\left(
k-k^{0}\right)  ,\label{DeltaE}%
\end{equation}
where $\sigma$ is the determinant of the unit 2-sphere. $k^{0}$ represents the
trace of the extrinsic curvature corresponding to embedding in the
two-dimensional boundary $^{2}S$ in three-dimensional Euclidean space at
infinity. In alternative to the ADM energy, we can use quasilocal energy to
compute such a difference, which is defined by Eq. (\ref{DeltaE}) but for a
finite two sphere. The main reason to use such a definition is that we can
extend the surface energy computation even to non-asymptoticaaly flat
spaces.\textbf{ }For this purpose, consider a manifold $\mathcal{M}$ composed
by two wedges $\mathcal{M}_{+}$ and $\mathcal{M}_{-}$, located in the right
and left sectors of a Kruskal diagram, respectively and bounded by two
three-dimensional disconnected timelike boundaries $B_{+}$ and $B_{-}$ located
in $\mathcal{M}_{+}$ and $\mathcal{M}_{-}$ respectively. The quasilocal energy
$E_{tot}$ of a spacelike hypersurface $\Sigma=\Sigma_{+}\cup\Sigma_{-}$
bounded by two spacelike boundaries $S_{+}$ and $S_{-}$ located in
$\mathcal{M}_{+}$ and $\mathcal{M}_{-}$ respectively, is given by
\cite{BrownYork,Frolov,Martinez}
\[
E_{tot}=E_{+}-E_{-}\,.
\]
More specifically, $E_{tot}$ is defined as the value of the Hamiltonian that
generates unit time translations orthogonal to the two-dimensional boundaries
\cite{BrownYork,Frolov,Martinez}. $E_{+}$ and $E_{-}$ are defined as
\begin{equation}
\left\{
\begin{array}
[c]{c}%
E_{+}=\frac{1}{8\pi G}\int_{S_{+}}d^{2}x\sqrt{\sigma}\left(  k-k^{0}\right)
\\
\\
E_{-}=-\frac{1}{8\pi G}\int_{S_{-}}d^{2}x\sqrt{\sigma}\left(  k-k^{0}\right)
\end{array}
\right.  ,\label{EBound}%
\end{equation}
respectively. The trace of the second fundamental form, $k$, is defined as%
\begin{equation}
k=-\frac{1}{\sqrt{h}}\left(  \sqrt{h}\,n^{\mu}\right)  _{,\mu},\label{trace}%
\end{equation}
where $n^{\mu}$ is the normal to the boundaries, and $h$ is the determinant of
the metric of $\Sigma$. As an example, consider the static Einstein-Rosen
bridge, with the metric given by%
\begin{equation}
ds^{2}=-N^{2}dt^{2}+g_{yy}dy^{2}+r^{2}\left(  y\right)  d\Omega^{2}%
,\label{a1b}%
\end{equation}
where the lapse function $N$, $g_{yy}$ and $r$ are functions of the radial
coordinate $y$ continuously defined on $\mathcal{M}$, with $dy=dr/\sqrt
{1-2MG/r}$. The boundaries $S_{+}$ and $S_{-}$ are located at coordinate
values $y=y_{+}$ and $y=y_{-}$, respectively, and the lapse function is given
by $\left\vert N\right\vert =1$ at both $S_{+}$ and $S_{-}$. In this case
$n^{\mu}=\left(  g^{yy}\right)  ^{1/2}\delta_{y}^{\mu}$. Since this normal is
defined continuously along $\Sigma$, the value of $k$ depends on the function
$r,_{y}$, which is positive for $B_{+}$ and negative for $B_{-}$. See figure
\ref{Penrosediagram} for a Penrose-Carter diagram illustrating the boundary
locations in a Schwarzschild metric. 

From Eq. (\ref{trace}) and Eq. (\ref{a1b}), we obtain at either boundary that
\begin{equation}
k=-\frac{2r,_{y}}{r},
\end{equation}
where we have assumed that the function $r,_{y}$ is positive for $S_{+}$ and
negative for $S_{-}$. The trace associated with the subtraction term is taken
to be $k^{0}=-2/r$ for $B_{+}$ and $k^{0}=2/r$ for $B_{-}$. As an
illustration, consider the case when the boundary $B_{+}$ is located at
right-hand infinity $\left(  y_{+}=+\infty\right)  $ and the boundary $B_{-}$
is located at $y_{-}$, then
\begin{equation}
E_{tot}=M-r\left[  1-\left(  1-\frac{2MG}{r}\right)  ^{\frac{1}{2}}\right]  .
\label{a1}%
\end{equation}
It is easy to see that $E_{+}$ and $E_{-}$ tend individually to the ADM mass
$M$ when the boundaries $^{3}B_{+}$ and $^{3}B_{-}$ tend respectively to right
and left spatial infinity. It should be noted that the total energy is zero
for boundary conditions symmetric with respect to the bifurcation surface,
i.e.,%
\begin{equation}
E=E_{+}-E_{-}=M+\left(  -M\right)  =0. \label{a2}%
\end{equation}

Consider now the dark energy star of metric (\ref{ds metric}) and a wormhole
defined by the \textit{shape function }$b\left(  r\right)  $, with the
following difference%
\begin{equation}
\left\{
\begin{array}
[c]{l}%
\text{Dark energy star (DS)\thinspace, \qquad}m\left(  r\right)  \text{ with
}r\in\left[  0,+\infty\right) \\
\text{Wormhole (W)\thinspace, \qquad}b\left(  r\right)  \text{ with }%
r\in\left[  r_{0},+\infty\right)
\end{array}
\right.  .
\end{equation}

Consider also the relation $\left(  \ref{ADM:DS-W}\right)  $. Thus, by
repeating the computation leading to Eqs. (\ref{a1}) and (\ref{a2}) in the
case of interest, we get%
\begin{align}
k^{W}-k^{DS}  &  =\left(  k-k^{0}\right)  ^{W}-\left(  k-k^{0}\right)
^{DS}=\frac{-2}{r}\left(  r,_{y}-1\right)  ^{W}-\frac{-2}{r}\left(
r,_{y}-1\right)  ^{DS}\nonumber\\
&  =\frac{-2}{r}\left[  \sqrt{1-\frac{b\left(  r\right)  }{r}}-\sqrt
{1-\frac{2m\left(  r\right)  }{r}}\right]  , \label{Delta0}%
\end{align}
where we are looking at the positive wedge $\mathcal{M}_{+}$ only. For large
boundaries $R\gg r_{0}$ and expanding around the throat, one obtains%
\begin{equation}
k^{W}-k^{DS}=\frac{-2}{r}\left[  \sqrt{1-\frac{b\left(  r_{0}\right)
+b^{\prime}\left(  r_{0}\right)  \left(  r-r_{0}\right)  +\ldots}{r}}%
-\sqrt{1-\frac{2m\left(  r\right)  }{r}}\right]
\end{equation}%
\begin{align}
&  \simeq\frac{-2}{R}\left[  \left(  1-\frac{b\left(  r_{0}\right)
+b^{\prime}\left(  r_{0}\right)  \left(  r-r_{0}\right)  +\ldots}{2R}\right)
-\left(  1-\frac{m\left(  R\right)  }{R}\right)  \right] \nonumber\\
&  =\frac{1}{R^{2}}\left[  r_{0}+b^{\prime}\left(  r_{0}\right)  \left(
r-r_{0}\right)  -2m\left(  R\right)  \right]  ,
\end{align}
where we have used the wormhole condition at the throat, $b(r_{0})=r_{0}$. If
$b^{\prime}\left(  r_{0}\right)  =0$ and $m\left(  R\right)  $ is negligible,
then we recover the ADM mass. Indeed, by integrating on the boundary
$^{2}S_{+}$, we obtain%
\begin{equation}
E_{+}=\frac{r_{0}}{2G}=M.
\end{equation}
If $b^{\prime}\left(  r_{0}\right)  \neq0$ and $m\left(  R\right)  $ is not
vanishing, then the evaluation of the energy depends on a case to case
scenario. The same discussion can be applied on the negative wedge. As shown
in Eq. (\ref{a2}), if we choose boundaries symmetric with respect to the
bifurcation surface, here represented by the throat $r_{0}$, we have a total
zero ADM-like energy. The physical situation looks like a familiar QED
physical process $\gamma\rightarrow e^{+}e^{-}$: the electric charge is
conserved. In our case, the charge is the asymptotic energy. Since there is no
reason to have an asymmetry in boundaries in the absence of external forces,
we have to conclude that the classical term is not able to predict the
appearance of a wormhole or the permanence of a dark star. We are forced to
compute quantum effects. The implicit subtraction procedure of Eq.
(\ref{EBound}), can be extended in such a way that we can include quantum
effects: this is the Casimir energy or in other terms, the vacuum energy. One
can in general formally define the Casimir energy as follows%
\begin{equation}
E_{Casimir}\left[  \partial\mathcal{M}\right]  =E_{0}\left[  \partial
\mathcal{M}\right]  -E_{0}\left[  0\right]  , \label{a3}%
\end{equation}
where $E_{0}$ is the zero-point energy, $\partial\mathcal{M}$ is a boundary
and $E_{0}\left[  0\right]  $ represents the zero point energy without a
boundary. For zero temperature, the idea underlying the Casimir effect is to
compare vacuum energies in two physical distinct configurations. The extension
to quantum effects is straightforward
\begin{equation}
E_{Casimir}\left[  \partial\mathcal{M}\right]  =\left(  E_{0}\left[
\partial\mathcal{M}\right]  -E_{0}\left[  0\right]  \right)  _{classical}%
+\left(  E_{0}\left[  \partial\mathcal{M}\right]  -E_{0}\left[  0\right]
\right)  _{1-loop}+\ldots.
\end{equation}

In our picture, the classical part represented by the ADM-like energy is
vanishing, because of the symmetry of boundary conditions. This means that%
\begin{equation}
E_{Casimir}\left[  \partial\mathcal{M}\right]  =\left(  E_{0}\left[
\partial\mathcal{M}\right]  -E_{0}\left[  0\right]  \right)  _{1-loop}%
+\ldots.,
\end{equation}
namely $E_{Casimir}$ is purely quantum. Thus, the Casimir energy can be
regarded as a measure of the topology change. With this, we mean that, if
$E_{Casimir}$ is positive then the topology change will be suppressed, while
if it is negative, it will be favored. It is important to remark that in most
physical situations, the Casimir energy is negative. Consider now the one loop
term. We will evaluate it following the scheme of Eq. (\ref{Delta0}). Thus
\begin{equation}
\left(  E_{0}^{W}\left[  \partial\mathcal{M}\right]  -E_{0}^{DS}\left[
\partial\mathcal{M}\right]  \right)  _{1-loop}=\left(  E_{0}^{W}\left[
\partial\mathcal{M}\right]  -E_{0}\left[  0\right]  \right)  _{1-loop}+\left(
E_{0}\left[  0\right]  -E_{0}^{DS}\left[  \partial\mathcal{M}\right]  \right)
_{1-loop}\,. \label{Delta1}%
\end{equation}

The procedure followed to evaluate Eq. (\ref{Delta1}), relies heavily on the
formalism outlined in Refs. \cite{Remo,Remo1}. The computation was realized
through a variational approach with Gaussian trial wave functionals. A zeta
function regularization is used to deal with the divergences, and a
renormalization procedure is introduced, where the finite one loop is
considered as a self-consistent source for traversable wormholes. Rather than
reproduce the formalism, we shall refer the reader to Refs. \cite{Remo,Remo1}
for details, when necessary. We can write,%
\begin{align}
&  \left(  E_{0}^{W}\left[  \partial\mathcal{M}\right]  -E_{0}^{DS}\left[
\partial\mathcal{M}\right]  \right)  _{1-loop}=\frac{1}{64\pi^{2}}\left\{
\left[  \left(  m_{L}^{2}\left(  r\right)  +m_{1,S}^{2}\left(  r\right)
\right)  ^{2}\ln\left(  \frac{m_{L}^{2}\left(  r\right)  +m_{1,S}^{2}\left(
r\right)  }{4\mu_{0}^{2}}\sqrt{e}\right)  \right.  \right. \nonumber\\
&  \left.  +\left(  m_{L}^{2}\left(  r\right)  +m_{2,S}^{2}\left(  r\right)
\right)  ^{2}\ln\left(  \frac{m_{L}^{2}\left(  r\right)  +m_{2,S}^{2}\left(
r\right)  }{4\mu_{0}^{2}}\sqrt{e}\right)  \right]  _{W}-\left[  \left(
m_{L}^{2}\left(  r\right)  +m_{1,S}^{2}\left(  r\right)  \right)  ^{2}%
\ln\left(  \frac{m_{L}^{2}\left(  r\right)  +m_{1,S}^{2}\left(  r\right)
}{4\mu_{0}^{2}}\sqrt{e}\right)  \right. \nonumber\\
&  \hspace{3cm}\left.  \left.  +\left(  m_{L}^{2}\left(  r\right)
+m_{2,S}^{2}\left(  r\right)  \right)  ^{2}\ln\left(  \frac{m_{L}^{2}\left(
r\right)  +m_{2,S}^{2}\left(  r\right)  }{4\mu_{0}^{2}}\sqrt{e}\right)
\right]  _{DS}\right\}  , \label{Delta2}%
\end{align}
where we have defined two $r$-dependent effective masses $m_{1}^{2}\left(
r\right)  $ and $m_{2}^{2}(r)$, which can be cast in the following form%
\begin{equation}
\left\{
\begin{array}
[c]{c}%
m_{1}^{2}\left(  r\right)  =m_{L}^{2}\left(  r\right)  +m_{1,S}^{2}\left(
r\right) \\
\\
m_{2}^{2}\left(  r\right)  =m_{L}^{2}\left(  r\right)  +m_{2,S}^{2}\left(
r\right)
\end{array}
\right.  ,
\end{equation}
where%
\begin{equation}
m_{L}^{2}\left(  r\right)  =\frac{6}{r^{2}}\left(  1-\frac{b\left(  r\right)
}{r}\right)  \label{mL}%
\end{equation}
and%
\begin{equation}
\left\{
\begin{array}
[c]{c}%
m_{1,S}^{2}\left(  r\right)  =\left[  \frac{3}{2r^{2}}b^{\prime}\left(
r\right)  -\frac{3}{2r^{3}}b\left(  r\right)  \right] \\
\\
m_{2,S}^{2}\left(  r\right)  =\left[  \frac{1}{2r^{2}}b^{\prime}\left(
r\right)  +\frac{3}{2r^{3}}b\left(  r\right)  \right]
\end{array}
\right.  , \label{m12s}%
\end{equation}
respectively. We refer the reader to Refs. \cite{Remo,Remo1} for the deduction
of these expressions in the Schwarzschild case. The zeta function
regularization method has been used to determine the energy densities,
$\rho_{i}$. It is interesting to note that this method is identical to the
subtraction procedure of the Casimir energy computation, where the zero point
energy in different backgrounds with the same asymptotic properties is
involved. In this context, the additional mass parameter $\mu$ has been
introduced to restore the correct dimension for the regularized quantities.
Note that this arbitrary mass scale appears in any regularization scheme. Of
course $b\left(  r\right)  =2m\left(  r\right)  $, then we can use only one
function recalling the different boundary conditions they must satisfy.
Generally speaking we can adopt the condition $m\left(  0\right)  =0$ for the
dark energy star and $b\left(  r_{0}\right)  =r_{0}$ for the wormhole. Thus,
the leading part related to the dark energy star close to $r=0$, simply
becomes%
\[
-\left(  E_{0}\left[  0\right]  -E_{0}^{DS}\left[  \partial\mathcal{M}\right]
\right)  _{1-loop}\simeq\left[  -\frac{3}{16\pi^{2}r^{4}}\ln\left(  \frac
{6}{4r^{2}\mu_{0}^{2}}\sqrt{e}\right)  \right]  _{DS}.
\]
On the other hand for the wormhole we get at the throat \footnote{Actually in
Eq. (\ref{Delta1}), the argument of the log has an absolute value.}%
\begin{align}
\left(  E_{0}^{W}\left[  \partial\mathcal{M}\right]  -E_{0}\left[  0\right]
\right)  _{1-loop}  &  =\frac{1}{64\pi^{2}}\Bigg[\frac{9}{4r_{0}^{4}}\left(
b^{\prime}\left(  r_{0}\right)  -1\right)  ^{2}\ln\left(  \left\vert
\frac{b^{\prime}\left(  r_{0}\right)  -1}{8r_{0}^{2}\mu_{0}^{2}}\sqrt
{e}\right\vert \right) \nonumber\\
&  +\frac{1}{4r_{0}^{4}}\left(  b^{\prime}\left(  r_{0}\right)  +3\right)
^{2}\ln\left(  \frac{b^{\prime}\left(  r_{0}\right)  +3}{8r_{0}^{2}\mu_{0}%
^{2}}\sqrt{e}\right)  \Bigg]_{W}. \label{DeltaEW}%
\end{align}

To have an easy comparison with the dark energy star, we make a specific
choice for the wormhole shape function. We assume that%
\begin{equation}
b\left(  r\right)  =\frac{r_{0}^{2}}{r}, \label{shape}%
\end{equation}
then we obtain%
\begin{equation}
\left(  E_{0}^{W}\left[  \partial\mathcal{M}\right]  -E_{0}^{DS}\left[
\partial\mathcal{M}\right]  \right)  _{1-loop}\simeq\frac{1}{64\pi^{2}%
}\left\{  \frac{10}{r_{0}^{4}}\ln\left(  \frac{\sqrt{e}}{4r_{0}^{2}\mu_{0}%
^{2}}\right)  -\left[  \frac{12}{r^{4}}\ln\left(  \frac{6\sqrt{e}}{4r^{2}%
\mu_{0}^{2}}\right)  \right]  \right\}  .
\end{equation}
Moreover, we evaluate the dark energy star term close to $r_{0}$ to get%
\begin{equation}
\left(  E_{0}^{W}\left[  \partial\mathcal{M}\right]  -E_{0}^{DS}\left[
\partial\mathcal{M}\right]  \right)  _{1-loop}\simeq\frac{1}{32\pi^{2}%
r_{0}^{4}}\left\{  5\ln\left(  \frac{\sqrt{e}}{4r_{0}^{2}\mu_{0}^{2}}\right)
-\left[  6\ln\left(  \frac{6\sqrt{e}}{4r_{0}^{2}\mu_{0}^{2}}\right)  \right]
\right\}  .
\end{equation}
If we choose%
\begin{equation}
\mu_{0}\leq\frac{108\sqrt[4]{e}}{r_{0}}\qquad\Longrightarrow\qquad\left(
E_{0}^{W}\left[  \partial\mathcal{M}\right]  -E_{0}^{DS}\left[  \partial
\mathcal{M}\right]  \right)  _{1-loop}\leq0. \label{ineq}%
\end{equation}
It is important to remark that the result of inequality (\ref{ineq}) is valid
only for the class of traversable wormholes expressed by the shape function
(\ref{shape}). To discuss the appearance of different class of traversable
wormholes, we need to use expression (\ref{DeltaEW}) inside inequality
(\ref{ineq}) and it is quite evident that this strongly depends on the form of
the shape function as it should be. It is interesting to note that once this
has been created, there is a probability that it will be self-sustained
\cite{Garattini:2007ff}, at least for an inhomogeneous $\omega$ parameter,
like in our case. This means that quantum fluctuations related to the Casimir
energy play a fundamental part not only for the topology change but even for
the traversable wormhole persistence.


\subsection{Morse Index Analysis}

\begin{figure}[th]
\begin{center}
\vspace{0.5cm}
\includegraphics[bb=0 0 900 410, scale=0.4, clip, keepaspectratio=true]
{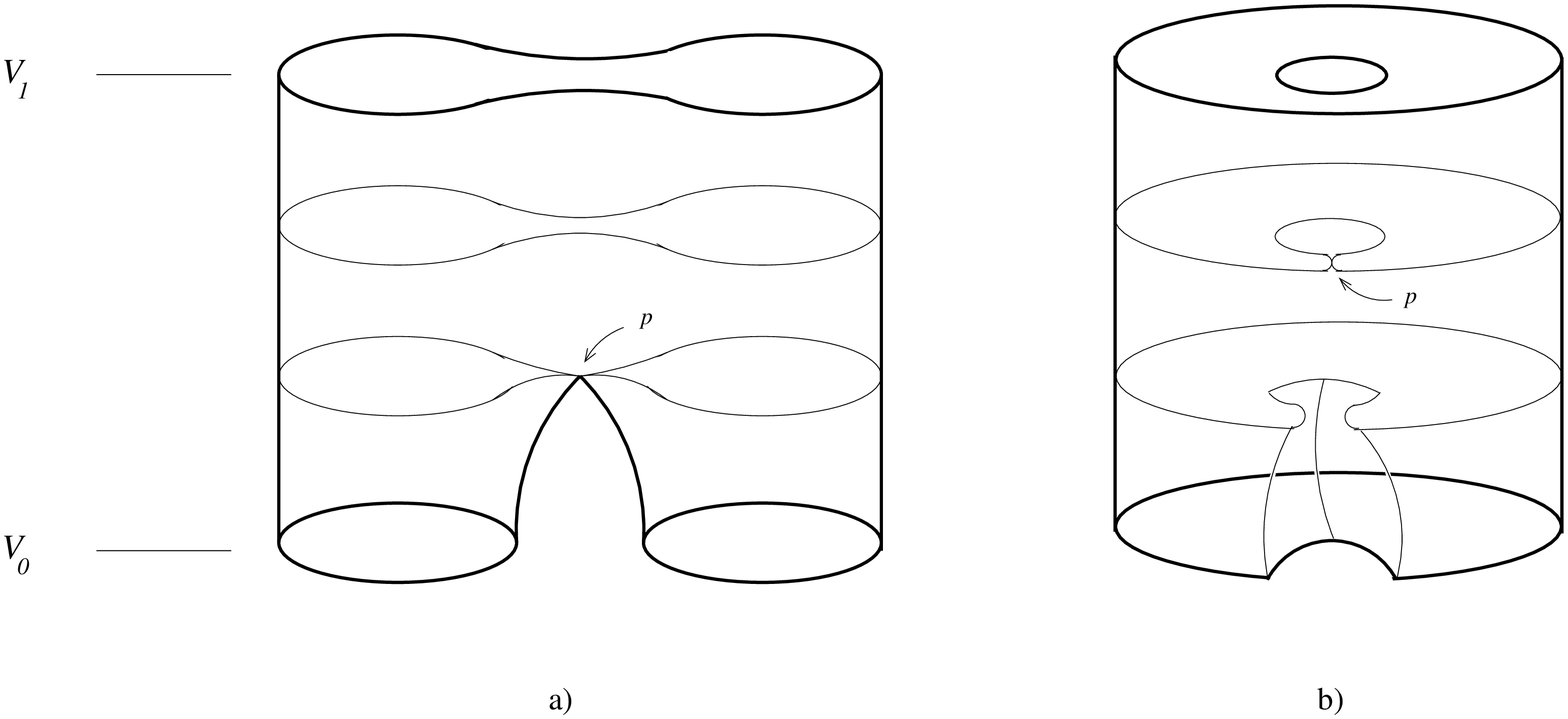}
\end{center}
\caption{{\small Schematics of topology changing space-times via wormhole
formation. Figure a) represents topology change via the formation of an
inter-universe wormhole. Figure b) represents topology change via the
formation of an intra-universe wormhole. The points $p$ represent the critical
point of the topology change.}}%
\label{fig:tc1}%
\end{figure}

In the classical case there are arguments that if $V_{0}$ and
$V_{1}$ are compact 3-manifolds, there will exist a space-time
whose boundary is comprised of the disjoint union of $V_{0}$ and
$V_{1}$ \cite{Morseindex} (see figure \ref{fig:tc1} for
reference.) In relation, Geroch's theorem states that if $V_{0}$
and $V_{1}$ possess differing topology, then a singularity or
closed time-like curves must exist somewhere on the manifold
\cite{Morseindex}. In the realm of wormhole physics one
has to generally accept the possibility of closed time-like curves
unless the kinematics of the wormhole are constrained in some
manner \cite{mty}. This is true regardless of whether there is
topology change or not and is simply a consequence of having a
wormhole whose mounths may move relative to each other. Therefore,
in this sense, the issue of closed time-like curves is no more
serious a problem in the topology changing scenarios than in
``standard'' wormhole physics.

Regarding the singularities, even if some singular behaviour exists
classically, one may argue that topology changing space-times may still
contribute to the Lorentzian functional integral approach to quantum gravity
where one considers the functional integral:
\begin{equation}
\label{eq:int}I=\int D[g] e^{iS} \,.
\end{equation}
Here $S$ is the usual Einstein-Hilbert action. It has been convincingly argued
that some of the singularities that arise in certain topology changing
space-times are extremely mild \cite{Morseindex} in the sense that the tetrad
becomes degenerate but the equations of motion (and the resulting curvature)
remain well defined. As well, the Loop quantum gravity approach relies on
(densitized) tetrads and self-dual connections, and it is know that solutions
with classically degenerate tetrads yield finite equations of motion also
using these Ashtekar variables. Therefore, classically degenerate tetrads are
not necessarily an Achille's heel in this theory. In fact, it is possible that
degenerate tetrads may play an important role in quantum gravity,
\cite{Morseindex}. 

Having established that some of the pathologies associated with topology
change are ``mild'', the natural question to ask is what type of pathology
accompanies various topology changes. In this respect, the picture is less
clear and many studies in the literature are based on case by case bases.
Horowitz \cite{Morseindex} has convincingly argued that by allowing the
possibility of degenerate tetrads, topology change is unavoidable.

Some studies regarding the feasibility of actual topology change rely on the
study of Morse functions on the topology changing space-times
\cite{Morseindex}. These studies indicate that manifolds with
critical points of Morse index 1 or $D-1$ ($D$ being the dimension of the
manifold) possess causal discontinuities of a severity which are problematic
in semi-classical analysis (the Borde-Sorkin conjecture \cite{Morseindex}).
In brief, on the topology changing manifold one constructs a Morse function on
the metric with $V_{0}$ and $V_{1}$ as boundaries \footnote{Readers unfamiliar
with Morse functions and the Morse index are referred to \cite{ref:topbook}.
Roughly, the Morse index at $p$ measures the number of maxima of the Morse
function at $p$ or equivalently, the number of independent directions in which
the manifold is concave-down.}. The Morse function, denoted usually by $f$,
possesses critical points where $\partial_{\alpha}f=0$ for all values of
$\alpha$. The Morse index, $\lambda_{p}$ at the critical points measures the
number of negative eigenvalues possessed by the matrix $\partial_{\alpha
}\partial_{\beta}f$ at such points.

In the space-times with Morse index 1 or $D-1$, the causal discontinuity
causes the propagation of quantum scalar fields to become singular somewhere
on the manifold, at least in $1+1$ dimensions \cite{Morseindex}.
This has been used to argue that these type of topology changing space-times
are suppressed in the sum over manifolds in (\ref{eq:int}), due to the fact
that they are highly sensitive to small fluctuations. Therefore, such metrics
would not make a significant contribution to the sum over manifolds due to
their combined destructive interference. If this were the case, such topology
changing spacetimes would be unlikely.

Figure \ref{fig:tc1} illustrates a dimensionally reduced schematic of two
types of wormhole formation, specifically an inter-universe and intra-universe
wormhole formation. The critical points of the topology change are denoted by
$p$. In such scenarios, the Morse index at $p$ is 1 and such topology change
would be suppressed according to the above argument in the sum over manifolds
approach to quantum gravity. Physically, the causal discontinuity occurs
because at the onset of wormhole formation, at least two points that were
previously not in causal contact suddenly become causally connected. (The key
point is that this is due to the topology change as opposed to the usual
``passage of time''.) However we hasten to add that at the moment it is not
clear that such topology changes are completely forbidden, even in the sum
over manifolds approach, keeping in mind that small probability is quite
different than no probability. It is also unknown if the sum over Lorentzian
manifolds approach to quantum gravity is indeed a valid method to calculate
probabilities in a quantum theory of gravity. Although there are now promising
candidate theories of quantum gravity, it is unknown which, if any, provide
the correct methods for calculating properties of quantum space-time.


\section{Summary and discussion}

\label{sec:conclusion}

In this work, we have considered time-dependent dark energy star models, with
an evolving parameter $\omega$ crossing the phantom divide, $\omega=-1$. In
particular, we briefly reviewed static and spherically symmetric dark energy
stars, and further analyzed general solutions of time-dependent spacetimes in
detail. Specific time-dependent solutions were extensively explored, in
particular, the specific cases of a constant energy density, the
Tolman-Matese-Whitman mass function solution, and a class of models with a
non-zero energy flux term, which form from gravitational collapse. Once the
parameter $\omega$ evolves into the phantom regime, the null energy condition
is violated, which physically implies that the negative radial pressure
exceeds the energy density. Therefore, an enormous negative pressure at the
center may, in principle, imply a topology change, consequently opening up a
tunnel and converting the dark energy star into a wormhole. The theoretical
difficulties and criteria for this topology change were discussed in detail,
where in particular we considered a Casimir energy approach involving
quasi-local energy difference calculations that may reflect or measure the
occurrence of a topology change. Once the topology change has occurred, it is
possible that the resulting wormhole structures, supported by phantom energy,
be self-sustained. As mentioned in the Introduction, recent fits to
observational data probably favor an evolving equation of state, with the dark
energy parameter crossing the phantom divide $\omega=-1$ \cite{Vikman}.
However, in a cosmological setting the transition into the phantom regime is
physically implausible for a single scalar field \cite{Vikman}, so that a
possible approach would be to consider a mixture of interacting non-ideal
fluids. One may consider that the time-dependent dark energy star model
outlined in this work, is a simplification of this possible approach. In fact,
recently, static models with two interacting phantom and ghost scalar fields
were considered, and it was shown that regular solutions exist
\cite{Dzhunushaliev:2007cs}. It would be interesting to generalize the latter
study to time-dependent solutions, extending the analysis considered in this work.

It is interesting to note that the topology change at the center should
influence the surface stresses at the thin shell, as there is a redistribution
of the stress-energy tensor components of the interior solution during the
change in topology. That this is so may be verified through the conservation
identity given by $S^{i}_{j|i}=[T_{\mu\nu}e^{\mu}_{(j)}n^{\nu}]^{+}_{-}\,$,
where $[X]^{+}_{-}$ denotes the discontinuity across the surface interface
$\Sigma$, i.e., $[X]^{+}_{-}=X^{+}|_{\Sigma}-X^{-}|_{\Sigma}$. The quantity
$S^{i}_{j}$ is the surface stress-energy tensor at the junction surface
$\Sigma$; $n^{\mu}$ is the unit normal $4-$vector to $\Sigma$; and $e^{\mu
}_{(i)}$ are the components of the holonomic basis vectors tangent to $\Sigma$
(see Refs. \cite{WHshell} for details). Note the dependency of the
conservation identity on the stress-energy tensor $T_{\mu\nu}$, and the right
hand side of the conservation identity may also be written as $S^{i}_{\tau
|i}=-\left[  \dot{\sigma}+2\dot{a}(\sigma+\mathcal{P} )/a \right]  $. The
momentum flux term, i.e., $[T_{\mu\nu}e^{\mu}_{(j)}n^{\nu}]^{+}_{-}\,$,
corresponds to the net discontinuity in the momentum flux $F_{\mu}=T_{\mu\nu
}\,U^{\nu}$ which impinges on the shell. The conservation identity is a
statement that all energy and momentum that plunges into the thin shell, gets
caught by the latter and converts into conserved energy and momentum of the
surface stresses of the junction. Now, it may be that the topology change is
sufficiently violent to disrupture the thin shell. On the other hand, one may
also assume that it is sufficiently mild as not to significantly affect the
stability of the surface layer.

In analogy to the case outlined in Ref. \cite{Roman:1992xj}, where the
possibility that inflation might provide a natural mechanism for the
enlargement of wormholes to macroscopic size was explored, one could imagine
that microscopic wormholes originated through a topology change, and due to
the accelerated expansion of the Universe, these submicroscopic constructions
could naturally be grown to macroscopic dimensions. For instance, in Ref.
\cite{gonzalez2} the evolution of wormholes and ringholes embedded in a
background accelerating Universe driven by dark energy, was analyzed. It was
shown that the wormhole's size increases by a factor proportional to the scale
factor of the Universe, and still increases significantly if the cosmic
expansion is driven by phantom energy. The accretion of dark and phantom
energy onto Morris-Thorne wormholes~\cite{diaz-phantom3,diaz-phantom4}, was
further explored, and it was shown that this accretion gradually increases the
wormhole throat which eventually overtakes the accelerated expansion of the
universe, consequently engulfing the entire Universe, and becomes infinite at
a time in the future before the Big Rip. This process was dubbed the ``Big
Trip'' \cite{diaz-phantom3,diaz-phantom4}. However, in the context of the
generalized Chaplygin gas, it was shown that the Big Rip may be avoided
altogether \cite{diaz-phantom,Madrid}. We refer the reader to Refs.
\cite{Yurov:2006we} for more recent details on these issues. In summary, we
denote these exotic geometries consisting of dark energy stars (in the phantom
regime) and phantom wormholes as \textit{phantom stars}. The final product of
this topological change, namely, phantom wormholes, have far-reaching physical
and cosmological implications, as in addition to being used for interstellar
shortcuts, an absurdly advanced civilization may manipulate these geometries
to induce closed timelike curves, consequently violating causality.

Relative to the topology change issue, a few words are in order. We emphasize
that it is still uncertain how to obtain this change in topology, and if
possible, it is riddled with technical and physical difficulties, such as the
appearance of closed timelike curves. Nevertheless, it is likely that enormous
negative pressures at the center is accompanied by a large quantum fluctuation
of the metric. The geometry of spacetime undergoing quantum fluctuations does
not seem to be a source of disagreement in the literature, but the question of
whether or not the topology of spacetime undergoes quantum fluctuations is
more subtle. In the latter, Wheeler conjectured that spacetime could be
subjected to a topology fluctuation at the Planck scale, where spacetime
undergoes a deep and rapid transformation in its structure, resulting in a
spacetime quantum foam, which can be taken as a model for the quantum
gravitational vacuum. Nevertheless, how to realize such a foam-like space and
also whether this represents the real quantum gravitational vacuum are still
unknown. From the quantum point of view we can separate the problem of
topology change generated by a canonical quantization approach and a
functional integral quantization approach. As mentioned above, the Hawking
topology change theorem is thus enough to show that the topology of space
cannot change in canonically quantized gravity. In the Feynman functional
integral quantization of gravitation things are different, where an approach
to spacetime foam is possible where fluctuations of topology become an
important phenomenon at least at the Planck scale. 

\acknowledgments FSNL was funded by Funda\c{c}\~{a}o para a Ci\^{e}ncia e a
Tecnologia (FCT)--Portugal through the grant SFRH/BPD/26269/2006.


\end{document}